\documentclass[11pt, reqno]{article}
\usepackage{jheppub}
\usepackage{diagbox}
\usepackage[usenames,dvipsnames]{xcolor}
\usepackage{amsmath,amssymb,mathrsfs,amsthm,tikz,shuffle,paralist}
\usepackage{dsfont}
\usepackage{relsize}
\definecolor{darkred}{RGB}{173,34,48}
\usepackage[all]{xypic}
\usepackage{subfigure}
\usepackage{caption}
\usepackage{graphicx}
\usepackage{extarrows}
\usepackage{mathtools}
\usepackage{epsfig}
\usepackage{subfigure,tikz}
\usetikzlibrary{backgrounds,automata}
\usepackage{graphics}
\usepackage[active]{srcltx}
\usepackage[T1]{fontenc}
\newcommand{\bea}{\begin{eqnarray}}
\newcommand{\eea}{\end{eqnarray}}
\newcommand{\bean}{\begin{eqnarray*}}
	\newcommand{\eean}{\end{eqnarray*}}
\newcommand{\nn}{\nonumber \\}

\def\lowerterm{\text{Known\ Terms}}

\def\D #1{\dot{#1}}
\def\W #1{\widetilde{#1}}

\def\eref#1{(\ref{#1})}

\def\d{{\rm d}}
\def\wt{\widetilde}
\def\intll{\int_{1,2}}

\def\a{{\alpha}}

\def\d{\partial}

\def\eps{\epsilon}

\def\newline{{\hspace{15pt}}}

\def\what{\widehat}

\def\co{\,,}
\def\ed{\,.}

\newcommand{\intg}[2]{{\cal I}[#1,#2]}
\newcommand{\Rterm}[3]{{\cal R}_{\ell_{#3},Tad}^{(#1,#2)}}
\newcommand{\hatintg}[3]{{\cal I}[#1,#2;\what{#3}]}

\def\label#1{\label{#1}%
	\smash{\hbox to0pt{\raise1ex\hbox{\tiny[#1]}\hss}}}

\title{\boldmath PV-Reduction of Sunset Topology with Auxiliary Vector}

\author[a,b,c,d]{Bo Feng,}
\author[b,1]{Tingfei Li\note{Corresponding author.}}
\affiliation[a]{ Beijing Computational Science Research Center, Beijing 100084, China}
\affiliation[b]{Zhejiang Institute of Modern Physics, Zhejiang University, Hangzhou, 310027, P. R. China }
\affiliation[c]{Center of Mathematical Science, Zhejiang University, Hangzhou, 310027, P. R. China}
\affiliation[d]{Peng Huanwu Center for Fundamental Theory, Hefei, Anhui, 230026, China}

\emailAdd{fengbo@zju.edu.cn}
\emailAdd{tfli@zju.edu.cn}

\abstract{Passarino-Veltman (PV) reduction method has been proved to be very useful for the computation of general one-loop integrals. However, not much progress has been made when applying to higher loops. Recently, we have improved the PV-reduction method by introducing an auxiliary vector. In this paper, we apply our new method to the simplest two-loop integrals, i.e., the sunset topology. We show how to use differential operators to establish algebraic recursion relations for reduction coefficients. Our algorithm can be easily applied to the reduction of integrals with arbitrary high-rank tensor structures. We demonstrate the efficiency of our algorithm by computing the reduction with the total tensor rank up to four.
}

\keywords{PV-reduction, Sunset Topology, Auxiliary Vector}

\begin{document}
	\maketitle
	\flushbottom
\section{Introduction}
Calculation of Feynman integrals at the multi-loop level is of great importance for both perturbation quantum field theory and particle experiments. The general strategy is to expand arbitrary integrals to finite simpler integrals (called Master Integrals) and then do the integration of master integrals only. Thus, tensor reduction of Feynman integrals is one of the critical steps for various realistic calculations in the Standard Model. There is a large body of research for one-loop reduction~\cite{ Brown:1952eu,Melrose:1965kb,Passarino:1978jh,tHooft:1978jhc,vanNeerven:1983vr, Stuart:1987tt,vanOldenborgh:1989wn,Bern:1992em,Bern:1993kr, Bern:1994cg,Fleischer:1999hq,Binoth:1999sp,Denner:2002ii,Duplancic:2003tv,Britto:2004nc,Britto:2005ha,Denner:2005nn,Anastasiou:2006jv,Anastasiou:2006gt,Ellis:2007qk,Ossola_2007,Ossola:2007bb,Forde:2007mi,Giele:2008ve,Ellis:2008ir}. For  reducing multi-loop integrals,
various ideas and methods have been developed. There are some works on two-loop tensor reduction for some special integrals \cite{Weiglein:1993hd,Usyukina:1994eg,Davydychev:1995mg,Davydychev:1995nq,Fleischer:1994ef,Passarino:2001wv}, while  algorithm for the reduction of general Feynman integrals  was proposed in \cite{Tarasov:1996bz,Tarasov:1998nx,Ferroglia:2003yj,Kotikov:2018wxe,Tkachov:1981wb,Chetyrkin:1981qh}.
Among these methods, the Integration-By-Parts (IBP) method \cite{Tkachov:1981wb,Chetyrkin:1981qh} is widely used and
has been implemented by some powerful programs such as {\sc FIRE}, {\sc LiteRed},{\sc Kira}, etc \cite{Smirnov:2008iw,Smirnov:2014hma,Lee:2013mka,Smirnov:2019qkx,vonManteuffel:2012np,Maierhofer:2017gsa,Gerlach:2022qnc,Liu:2022chg}. Although its popularity, the number of IBP identities grows very fast for the increasing number of mass scalars and higher rank of tensor structure, it will become a bottleneck when dealing with complicated physical processes. Even at the one-loop level, reducing a general tensor massive pentagon with the IBP method is difficult. Thus it is always welcome to find more efficient
reduction method.

PV-reduction \cite{Passarino:1978jh} for one-loop integrals has played a significant role in general one-loop computations in history. However, when trying to generalize the method to higher loops, it becomes hard, and there are not many works in the literature. In \cite{Weiglein:1993hd,Actis:2004bp}, the tensor two-loop integrals with only two external legs are reduced to scalar integrals with PV-reduction method. However, an extra massless propagator appears. Integrals with new topologies have been introduced as master integrals to address the problem. This results against the intuitive picture: Reduction is achieved by removing one or more propagators from the original topology. Thus the reduction and master integrals should be constrained to the original topology and its sub-topologies.

Recently, we have generalized the original PV-reduction method for one-loop integrals by introducing an auxiliary vector $R$ in \cite{Feng:2021enk,Hu:2021nia}. Our improved PV-reduction method can be easily carried out and compute reduction coefficients analytically by employing algebraic recursion relations. With these analytical expressions, one can do many things. For example, by taking limits properly, we can solve the reduction with propagators having general higher powers \cite{Feng:2022uqp}. When external momenta are not general, i.e., the Gram determinant becomes zero, master integrals in the basis will not be independent anymore. We can systematically study their degeneration \cite{Feng:2022rfz}.

Encouraged by the results in \cite{Feng:2021enk,Hu:2021nia}, we want to
see if such an improved PV-reduction method can be transplanted to higher loops. When moving to two loops, some complexities arise. The first is that we need to determine the master integrals. For one-loop, the master integrals are trivially known. For higher loops, the choice of master integrals becomes an art. With proper choice, one can reduce the computation greatly \cite{Henn:2013pwa}. The second is that when intuitively generalizing our method to two loops, we will meet the irreducible scalar products, which can not be reduced. In fact, these two complexities are related to each other. To avoid unnecessary complexity, we will focus on the simplest nontrivial two-loop integrals, i.e., the sunset topology in this paper. With this example, we will show how to generalize our new PV-reduction method to two loops.

This paper is structured as follows. In section \ref{sec:one-loop--two-loop}, we briefly review our previous work on one-loop tensor reduction and discuss the main idea for the reduction of sunset integrals. In section \ref{sec:recursions}, we derive recursion relations for sunset integrals using differential operators and the proper choice of the master integrals in our algorithm. In section \ref{sec:examples}, we demonstrate our algorithm successfully to sunset integrals with the total rank from one to four. In section \ref{sec:conclusion}, we give some discussions and the plan for further studying. Technical details are collected in Appendix.

\section{Brief discussion of tensor reduction with auxiliary vector}
\label{sec:one-loop--two-loop}

In this section, we will give a brief discussion of the general idea of using the auxiliary vector $R$ to do tensor reduction.
We first review the method presented in \cite{Feng:2021enk,Hu:2021nia}
and then move to the two-loop sunset integrals.
For the tensor reduction of a general one-loop integral (to avoid confusion, we denote $P_j$ as propagators for one-loop integrals while $D_j$ for two-loop integrals)
\bea
I_{n+1}^{\mu_1\mu_2\cdots\mu_m}\equiv \int {d^D \ell\over i\pi^{D/2}}  {\ell^{\mu_1}\ell^{\mu_2}... \ell^{\mu_m}\over P_0\prod_{j=1}^{n-1} P_j}=\int {d^D \ell\over i\pi^{D/2}}  {\ell^{\mu_1}\ell^{\mu_2}... \ell^{\mu_m}\over (\ell^2-M_0^2)\prod_{j=1}^{n} ((\ell-K_j)^2-M_j^2)},~~~~\label{set-1-1}
\eea
one can recover its tensor structure by multiplying each index with an auxiliary vector $R_{i,\mu_i}$. Furthermore, we can combine these $R_i$ to $R=\sum_{i=1}^m x_i R_i$ to simplify the expression \eref{set-1-1} to
\bea I^{(m)}_{n+1}\equiv\int {d^D \ell\over i\pi^{D/2}}  {(2R\cdot \ell)^m\over P_0\prod_{j=1}^{n} P_j}\ed~~~~\label{set-1-2}
\eea
The good point using \eref{set-1-2} instead of \eref{set-1-1} is that
we have avoided the complicated Lorentz tensor structure in the reduction process, while they can easily be retained
using the expansion of $R$ and taking corresponding coefficients of $x_i$'s.
Since we are using dimensional regularization, we like to keep the $D$ as a general parameter.
For $D=d-2\eps$, the master basis is the scalar integrals with  $n\leq d+1$ propagators, i.e., the  reduction results can be written as
\bea
I_{n}^{(m)}={ C}^{(m)}_{n \to n}I_{n}+{ C}^{(m)}_{n \to n;\what{i}}I_{n;\what{i}}+{ C}^{(m)}_{n \to n;\what{ij}}I_{n;\what{ij}}+\ldots
\eea
where we denote $I_{n;\what{i_1,\ldots,i_a}}$ as the integrals got by removing propagators $P_{i_1},P_{i_2},\ldots,P_{i_a}$ from  $I_n$. For simplicity, we denote
\begin{align}
s_{ij}\equiv K_i\cdot K_j,~~~s_{0i}\equiv R\cdot K_i,~~~f_i\equiv M_0^2-M_i^2+s_{ii}\ed
\end{align}

By the explicit permutation symmetry in \eref{set-1-1},
we only need to calculate  ${C}^{(m)}_{n+1 \to \{0,1,\ldots,r\}}\equiv {C}^{(m)}_{n+1 \to n+1; \what{r+1},\what{r+2},\ldots,\what{n}}$ while other reduction coefficients can be got by proper replacements  and momentum shifting. With the introduction of auxiliary vector $R$ in \eref{set-1-2},  one can expand the reduction coefficients according to their tensor structure
\begin{align}
C^{(m)}_{n+1 \to \{0,1,\ldots,r\}}
=& \sum_{2a_0+\sum_{k=1}^na_k=m}\bigg\{ c^{(0,1,\cdots,r)}_{a_1,\cdots,a_{n}}(m) (M_0^2)^{a_0+r-n}\prod_{k=0}^n s_{0k}^{a_k}\bigg\} \ed\label{coefficient-expansion-1}
\end{align}
By acting two types of differential operators $\mathcal{D}_i\equiv K_i\cdot {\d \over \d R} $ and $\cal{T}\equiv \eta^{\mu\nu}{\d\over \d R^{\mu}}{\d\over \d R^{\nu}}$ on \eref{set-1-2}, we get the recursion relations for the \textit{expansion coefficients} $c^{(0,1,\cdots,r)}_{a_1,\cdots,a_{n}}$
\begin{align}
\boldsymbol{c}^{(0,1,\cdots,r)}(a_1,\cdots,a_n;m)=\boldsymbol{T}^{-1} \widetilde{\boldsymbol{G}}^{-1}  \boldsymbol{O}^{(0,1,\cdots,r)}(a_1,\cdots,a_n;m)\co\label{inverse_formula}
\end{align}
\begin{align}
c^{(0,1,\cdots,r)}_{\underbrace{0,\cdots,0}_{\text{n\ times}}}(2k)
={2k-1\over  D+2 k-n-2}\bigg[(4-\boldsymbol{\a}^T\widetilde{\boldsymbol{G}}^{-1}\boldsymbol{\a})c_{\underbrace{0,\cdots,0}_{\text{n\ times}}}^{(0,1,\cdots,r)}(2k-2)+\boldsymbol{\a}^T\widetilde{\boldsymbol{G}}^{-1}\boldsymbol{c}_{\underbrace{0,\cdots,0}_{\text{n-1\ times}}}^{(0,1,\cdots,r)}(2k-2)\bigg],\label{Relations}
\end{align}
where $\widetilde{\boldsymbol{G}}=\left\{{s_{ij}\over M_0^2}\right\}$ is the $n\times n$ rescaled Gram matrix and  $\boldsymbol{T}=\text{diag}(a_1+1,a_2+1,\cdots,a_n+1)$ is a diagonal matrix. The vectors in the equations \eref{inverse_formula} and \eref{Relations} are defined as
\begin{align}
&\boldsymbol{\a}\equiv \left(\frac{f_1}{M_0^2},\frac{f_2}{M_0^2},\cdots,\frac{f_n}{M^2_0}\right)
\co ~ [\boldsymbol{c}^{(0,1,\cdots,r)}(a_1,\cdots,a_n;m)]_i\equiv  c^{(0,1,\cdots,r)}_{a_1,a_2,\cdots,a_i+1,\cdots, a_n}(m)\co
\end{align}
and
%
	\begin{align}
	[\boldsymbol{O}^{(0,1,\cdots,r)}(a_1,\cdots,a_n;m)]_i
	&\equiv  m\a_i c^{(0,1,\cdots,r)}_{a_1,\cdots,a_n}(m-1)- m  \delta_{0 a_i}  c^{(0,1,\cdots,r)}_{a_1,\cdots, \widehat{a}_i,\cdots, a_n}(m-1;\widehat{i}) \notag\\
	&\newline-(m+1-\sum_{l=1}^n a_l) c^{(0,1,\cdots,r)}_{a_1,\cdots,a_i-1,a_i,\cdots,a_n}(m)\co\nn
	\boldsymbol{c}_{\underbrace{0,\cdots,0}_{\text{n-1\ times}}}^{(0,1,\cdots,r)}(m)&\equiv
	\left(0,0,\cdots,0,c_{\underbrace{0,\cdots,0}_{\text{n-1\ times}}}^{(0,1,\cdots,r)}(m;\widehat{r+1}),\cdots,c_{\underbrace{0,\cdots,0}_{\text{n-1\ times}}}^{(0,1,\cdots,r)}(m;\widehat{n})\right)\ed
	\end{align}
%
where $\what{a}_i$ indicates to drop the index $a_i$. With the known boundary conditions, one can obtain all reduction coefficients by applying the recursions \eref{inverse_formula} and \eref{Relations}  iteratively.
\begin{table}
	\centering
	\begin{tabular}{>{\centering\arraybackslash}p{1.9cm}|>{\centering}p{1.9cm}>{\centering}p{1.9cm}>{\centering}p{1.9cm}>{\centering\arraybackslash}p{1.9cm}}
		\hline
		$\text{rank} \ m$ & $N_{c}$ & $N_{\cal D}$ & $N_{{\cal T}\cup{\cal D}}$ &$N_{c}-N_{{\cal T}\cup{\cal D}}$ \\
		\hline
		1 & 1 & 1 & 1 & 0 \\
		2 & 2 & 1 & 2 & 0 \\
		3 & 2 & 2 & 2 & 0 \\
		4 & 3 & 2 & 3 & 0 \\
		5 & 3 & 3 & 3 & 0 \\
		6&  4 & 3 & 4 & 0 \\
		\hline
	\end{tabular}
	\caption{Number of expansion coefficients and independent  equations for tensor bubble.}
	\label{bubble-DTeqnumbers}
\end{table}

One important point of above reduction method is that we need to use
both ${\cal D}_i$ and ${\cal T}$ operators to completely fix unknown
coefficients in \eref{coefficient-expansion-1}. One simple explanation
is that in \eref{coefficient-expansion-1} coefficients there are  $(n+1)$ indices ($a_i,i=1,...,n$ and rank $m$), so naively $(n+1)$  relations are needed. More accurate explanation is a little tricky. Let us take
the bubble, i.e., $n=1$ as an example. As shown in Table \underline{\ref{bubble-DTeqnumbers}}, for rank $m$, there are $N_{c}=(\lfloor{m\over 2}\rfloor+1)$ unknown coefficients while the number of independent $\cal D$-type equations is $N_{\cal D}=\lfloor {m+1\over 2}\rfloor$. Thus, only using the ${\cal D}$-type operators we can fix all
coefficients with odd rank $m=(2k+1)$ by these known terms with lower rank $m\leq 2k$. However, for $m=2k+2$, $N_{c}-N_{\cal D}=1$, thus just using
${\cal D}$-type relations is not enough and we need to adopt the ${\cal T}$ operator to provide one
extra independent relation. In fact, ${\cal T}$ provides more than just
one relation, but other relations are not independent to these coming
from ${\cal D}$ and can be taken as the consistent check of the reduction
method.

In this paper, we want to generalize our reduction method from one-loop to two-loop integrals. Again, with simplified tensor structure,  we define the general tensor integrals of sunset topology as\footnote{For simplicity, we denote the scalar integral $I_{a_1,a_2,a_3}\equiv I^{(0,0)}_{a_1,a_2,a_3}$.}
\bea
I_{a_1,a_2,a_3}^{(r_1,r_2)}\equiv \int {d^D\ell_1\over i\pi^{D/2}}{d^D\ell_2\over i\pi^{D/2}} {(2\ell_1\cdot R_1)^{r_1}(2\ell_2\cdot R_2)^{r_2}\over D_1^{a_1}D_2^{a_2}D_3^{a_3}}\label{tensor-sunset-def}
\eea
where the propagators are
\bea
D_1\equiv \ell_1^2-M_1^2\co~~~D_2\equiv \ell_2^2-M_2^2\co~~~~D_3\equiv(\ell_1+\ell_2-K)^2-M_3^2\,.~~~\label{3-D}
\eea
In this paper we consider only the reduction  with all $a_i=1$. For
$a_i\geq 2$, one can use the same strategy presented in \cite{Feng:2022uqp}.
For simplicity, we denote
\begin{align}
\int d \ell_{i}(\bullet ) &\equiv\int {d^D\ell_i\over i\pi^{D/2}}(\bullet )\co~~~~\int d \ell_{1,2}(\bullet )\equiv \int {d^D\ell_1\over i\pi^{D/2}}{d^D\ell_2\over i\pi^{D/2}}(\bullet )=\int_{1,2}(\bullet)\,.
\end{align}
Similar to what we do for one-loop case,  we can expand reduction coefficients according to their tensor structure\footnote{We will discuss the choice of master integrals ${\bf J}$ in section \ref{subsec:MIs}.}
\bea
I_{1,1,1}^{(r_1,r_2)}={\bf C}^{(r_1,r_2)}{\bf J}=\sum'_{i_1,i_2,j} s_{01}^{i_1}s_{0'1}^{i_2}s_{00'}^js_{00}^{r_1-i_1-j\over 2}s_{0'0'}^{r_2-i_2-j\over 2}{\vec{\a}}^{(r_1,r_2)}_{i_1,i_2,j}~{\bf J} \label{coeff-exp}
\eea
where we have written ${\bf C}$, ${\bf J}$ to emphasize they are
vector, while other kinematic variables are
\begin{align}
s_{01}=R_1\cdot K,~~ s_{0'1}=R_2\cdot K,~~ s_{11}=K^2,~~s_{00}=R_1^2,~~s_{0'0'}=R_2^2,~~~s_{00'}=R_1\cdot R_2\ed
\end{align}
In \eref{coeff-exp}, the \textit{expansion coefficients} ${\vec{\a}}^{(r_1,r_2)}_{i_1,i_2,j}$ are actually vectors with components corresponding to different master integrals $J_i$. The summation over indices $\{i_1,i_2,j\}\ge 0$ admits to  the restriction $0\le (r_l-i_l-j)/2\in \mathbb{N},l=1,2$. We make the indices free by setting all ${\vec{\a}}^{(r_1,r_2)}_{i_1,i_2,j}=0$ for inappropriate  indices throughout the paper, so we will drop the prime in other summations. Similar to the strategy  for one-loop case,  we solve expansion coefficients from lower rank levels to higher rank levels iteratively, where  the {\bf rank level} is defined as the  total rank $(r_1+r_2)$.

When we do the tensor reduction, we will reach the sub topologies where
one of the propagators in \eref{3-D} is removed. For these cases, the
integrals are reduced to the product of two one-loop tadpoles. Their
tensor reduction has been solved in Appendix \ref{lower-toplogy}. To simplify notation, we denote
\bea
I_{1,1,1; \what{i}}^{(r_1,r_2)}\equiv \int {d^D\ell_1\over i\pi^{D/2}}{d^D\ell_2\over i\pi^{D/2}} {(2\ell_1\cdot R_1)^{r_1}(2\ell_2\cdot R_2)^{r_2} D_i\over D_1 D_2D_3}\label{tensor-sunset-def-sub}
\eea
and their reductions are written as
\bea
I_{1,1,1;\what{i}}^{(r_1,r_2)}={\bf C}_{\what{i}}^{(r_1,r_2)}{\bf J}=\sum_{i_1,i_2,j} s_{01}^{i_1}s_{0'1}^{i_2}s_{00'}^js_{00}^{r_1-i_1-j\over 2}s_{0'0'}^{r_2-i_2-j\over 2}{\vec{\a}}^{(r_1,r_2)}_{i_1,i_2,j;\what{i}}~{\bf J}\ed \label{coeff-exp-sub}
\eea

In \eref{coeff-exp}, there are five indices for the expansion coefficients, i.e., $i_1, i_2, j,r_1, r_2$. With the experience from
one-loop integrals,  every index needs a recursion relation, so totally five kinds of differential operators are needed to give sufficient recursion relations. From the form of \eref{set-1-2} we can construct three $\mathcal{T}$-type operators
\bea
\mathcal{T}_{00}\equiv \eta^{\mu\nu}{\d \over \d R_1^{\mu}}{\d \over \d R_1^{\nu}}\co~~~
\mathcal{T}_{0'0'}\equiv \eta^{\mu\nu}{\d \over \d R_2^{\mu}}{\d \over \d R_2^{\nu}}\co~~~
\mathcal{T}_{00'}\equiv \eta^{\mu\nu}{\d \over \d R_1^{\mu}}{\d \over \d R_2^{\nu}}\ed~~~\label{T-def}
\eea
By applying these operators on the tensor integrals \eref{set-1-2}, we  produce combinations $\ell_1^2,\ell_2^2,\ell_1\cdot \ell_2$ in the numerator respectively. Using the following algebraic decomposition
\begin{align}
\ell_1^2&=D_1+M_1^2,~~~~~\ell_2^2=D_2^2+M_2^2\co\nn
2\ell_1\cdot \ell_2&=D_3+M_3^2-D_1-M_1^2-D_2-M_2^2-K^2+2\ell_1\cdot K+2\ell_2\cdot K~~~~\label{alg-rel-1}
\end{align}
we can write the result as a sum of  terms with one-lower or two-lower rank levels and terms for lower topologies, thus we can establish three recursion relations for these operators with expansion \eref{coeff-exp}.

Having discussed the operators ${\cal T}_i$, it is natural to think
about following two ${\cal D}$-type operators $K^{\mu}\cdot {\d \over \d R_1^{\mu}}$ and $K^{\mu}\cdot {\d \over \d R_2^{\mu}}$. However, when
acting on \eref{set-1-2}, it gives the combinations $\ell_1\cdot K$ and  $\ell_2\cdot K$. These two factors do not have simple algebraic
decompositions like \eref{alg-rel-1}. Thus it is not so easy to find
corresponding recursion relations. One key input in our paper
is that the recursion relation of ${\cal D}$-type operators can be established
if we consider the reduction of the sub one-loop integrals first.
As we will show, such a recursion relation is highly nontrivial and more
discussions will be given later.

\section{Recursion relations for tensor integrals of sunset topology}
\label{sec:recursions}
In this section, we first derive the recursion relations for expansion coefficients of three $\cal T$-type operators. Then we establish another two recursion relations of ${\cal D}$-type operators by reducing the left/right sub one-loop first. As we will see,
in the process, one needs to use a highly nontrivial relation of the reduction coefficients of tensor bubbles. Combining the recursion relations from ${\cal T}$-type and ${\cal D}$-type operators, we get five
relations. When using them to solve expansion coefficients, we find
that not all coefficients can be fixed. It indicates that some integrals with nontrivial numerators should be recognized as master integrals.

After introducing two auxiliary vectors $R_1, R_2$, we can define
differential operators (where we use the shorthand $\d_A={\d \over \d A}$)
\bea
{\d\over \d R_1^{\mu}}=2R_{1\mu}\d_{s_{00}}+K_{\mu}\d_{s_{01}}+R_{2\mu}\d_{s_{0'0}}\co~~{\d\over \d R_2^{\mu}}=2R_{2\mu}\d_{s_{0'0'}}+K_{\mu}\d_{s_{0'1}}+R_{1\mu}\d_{s_{0'0}}\ed~~~\label{R-def}
\eea
Using the two expressions and external Lorentz vectors, we can get following Lorentz
invariant combinations, i.e., three $\mathcal{T}$-type operators $\mathcal{T}_{00},\mathcal{T}_{00'},\mathcal{T}_{0'0'}$ defined in \eref{T-def} and six $\mathcal{D}$-type operators
\begin{align}
\mathcal{D}_{10}&\equiv K\cdot {\d\over \d R_1^{\mu}}\co\hspace{5pt} \mathcal{D}_{10'}\equiv K\cdot {\d\over \d R_2^{\mu}}\co\hspace{10pt}\mathcal{D}_{0'0}\equiv R_2\cdot {\d\over \d R_1^{\mu}}\co\nn
\mathcal{D}_{00'}&\equiv R_1\cdot {\d\over \d R_2^{\mu}}\co\hspace{5pt}\mathcal{D}_{00}\equiv R_1\cdot {\d\over \d R_1^{\mu}}\co\hspace{5pt}\mathcal{D}_{0'0'}\equiv R_2\cdot {\d\over \d R_2^{\mu}}\ed~~~\label{D-def}
\end{align}
It is easy to use the definition \eref{R-def} to find their action on reduction coefficients, for example,
\begin{align}
\mathcal{D}_{00'}&=2s_{0'0}\d_{s_{0'0'}}+s_{01}\d_{s_{0'1}}+s_{00}\d_{s_{00'}}\co\nn
\mathcal{T}_{00'}&=2\mathcal{D}_{00'}\d_{s_{00}}+\mathcal{D}_{10'}\d_{s_{01}}+D\d_{s_{00'}}+\mathcal{D}_{0'0'}\d_{s_{0'0}}\ed
\end{align}
Employing them, we can write down recursion relations for expansion
coefficients.

\subsection{Recursion relations from ${\cal T}$-type operators}
Here, we derive the recursion relations for three ${\cal T}$-type operators.
Let us start with the action of $\mathcal{T}_{00}$. It is easy to check
\bea
\mathcal{T}_{00}I_{1,1,1}^{(r_1\geq 2,r_2)}=4r_1(r_1-1)\left[I^{(r_1-2,r_2)}_{1,1,1;\what{1}}+M_1^2I^{(r_1-2,r_2)}_{1,1,1}\right]
\eea
by using the algebraic relation \eref{alg-rel-1}. Plugging the expansion
\eref{coeff-exp} into the both sides, we find
the LHS of the equation is
\begin{align}
&\sum_{i_1,i_2,j} \bigg[\left(r_1-i_1-j\right) \left(D+i_1+j+r_1-2\right){\vec{\a}}^{(r_1,r_2)}_{i_1,i_2,j}+2 (i_1+1)(j+1){\vec{\a}}^{(r_1,r_2)}_{i_1+1,i_2-1,j+1}\nn
&+\left(i_1+1\right) (i_1+2)s_{11}{\vec{\a}}^{(r_1,r_2)}_{i_1+2,i_2,j} +(j+1)(j+2){\vec{\a}}^{(r_1,r_2)}_{i_1,i_2,j+2}\bigg]s_{01}^{i_1}s_{0'1}^{i_2}s_{00'}^js_{00}^{r_1-2-i_1-j\over 2}s_{0'0'}^{r_2-i_2-j\over 2}{\bf J}\co
\end{align}
and the RHS is
\bea
4r_1(r_1-1)\sum_{i_1.i_2,j} \left({\vec{\a}}^{(r_1-2,r_2)}_{i_1,i_2,j;\what{1}}+M_1^2{\vec{\a}}^{(r_1-2,r_2)}_{i_1,i_2,j}\right)s_{01}^{i_1}s_{0'1}^{i_2}s_{00'}^js_{00}^{r_1-2-i_1-j\over 2}s_{0'0'}^{r_2-i_2-j\over 2}{\bf J}\ed
\eea
where the definition of ${\vec{\a}}^{(r_1-2,r_2)}_{i_1,i_2,j;\what{1}}$ has been given in \eref{coeff-exp-sub}.
Comparing both sides, we obtain  the recursion relation for operator $\mathcal{T}_{00}$ with $r_1\ge 2$
\begin{align}
&\left(i_1+1\right) (i_1+2)s_{11}{\vec{\a}}^{(r_1,r_2)}_{i_1+2,i_2,j}
+(j+1)(j+2){\vec{\a}}^{(r_1,r_2)}_{i_1,i_2,j+2}+2 (i_1+1)(j+1){\vec{\a}}^{(r_1,r_2)}_{i_1+1,i_2-1,j+1}\nn+&
\left(r_1-i_1-j\right) \left(D+i_1+j+r_1-2\right){\vec{\a}}^{(r_1,r_2)}_{i_1,i_2,j}=4r_1(r_1-1)\left [{\vec{\a}}^{(r_1-2,r_2)}_{i_1,i_2,j;\what{1}}+M_1^2{\vec{\a}}^{(r_1-2,r_2)}_{i_1,i_2,j}\right].\label{dR1dR1-Relation}
\end{align}
Similarly, one can derive the recursion relations for operator $\mathcal{T}_{0'0'}$ with $r_2\ge 2$
\begin{align}
&\left(i_2+1\right) (i_2+2)s_{11}{\vec{\a}}^{(r_1,r_2)}_{i_1,i_2+2,j}
+(j+1)(j+2){\vec{\a}}^{(r_1,r_2)}_{i_1,i_2,j+2}+2 (i_2+1)(j+1){\vec{\a}}^{(r_1,r_2)}_{i_1-1,i_2+1,j+1}\nn
+&\left(r_2-i_2-j\right) \left(D+i_2+j+r_2-2\right){\vec{\a}}^{(r_1,r_2)}_{i_1,i_2,j}=4r_2(r_2-1)\left[{\vec{\a}}^{(r_1,r_2-2)}_{i_1,i_2,j;\what{2}}+M_2^2{\vec{\a}}^{(r_1,r_2-2)}_{i_1,i_2,j}\right].\label{dR2dR2-Relation}
\end{align}
Finally, using the algebraic relation
\begin{align}
2\ell_1\cdot \ell_2
&=D_3-D_1-D_2+2\ell_1\cdot K+2\ell_2\cdot K-f_{12}
\end{align}
with
\bea f_{12}\equiv K^2+M_1^2+M_2^2-M_3^2\co~~~\label{f12}\eea
we derive recursion relations of operator $\mathcal{T}_{00'}$ for  $r_1\ge 1,r_2\ge 1$ as
\begin{align}
\mathcal{T}_{00'}I_{1,1,1}^{(r_1,r_2)}
=2r_1r_2\left[I^{(r_1-1,r_2-1)}_{1,1,1;\what{3}-\what{1}-\what{2}}-f_{12}I^{(r_1-1,r_2-1)}_{1,1,1}\right ]+2r_2{\mathcal{D}_{10}}I^{(r_1,r_2-1)}_{1,1,1}+2r_1{{\mathcal{D}_{10'}}}I^{(r_1-1,r_2)}_{1,1,1}\co \label{2l1l2-recur}
\end{align}
where we have used the shorthand
\begin{align}
I^{(r_1,r_2)}_{1,1,1;\what{3}-\what{1}-\what{2}}&\equiv I^{(r_1,r_2)}_{1,1,1;\what{3}}-I^{(r_1,r_2)}_{1,1,1;\what{1}}-I^{(r_1,r_2)}_{1,1,1;\what{2}}\nn
&=\sum_{i_1,i_2,j} s_{01}^{i_1}s_{0'1}^{i_2}s_{00'}^js_{00}^{r_1-i_1-j\over 2}s_{0'0'}^{r_2-i_2-j\over 2}{\vec{\a}}^{(r_1,r_2)}_{i_1,i_2,j;\what{3}-\what{1}-\what{2}}{\bf J}\ed
\end{align}
Employing the expansion \eref{coeff-exp}, we find the LHS of \eref{2l1l2-recur} is
\begin{align}
&\sum_{i_1,i_2,j}\Big[(j+1) \left(D+r_1+r_2-j-2\right){\vec{\a}}^{(r_1,r_2)}_{i_1,i_2,j+1}-(i_2+1)  \left(i_1-1+j-r_1\right){\vec{\a}}^{(r_1,r_2)}_{i_1-1,i_2+1,j}\nn
&\hspace{15pt}+\left(i_1+j-1-r_1\right) \left(i_2+j-1-r_2\right) {\vec{\a}}^{(r_1,r_2)}_{i_1,i_2,j-1}-(i_1+1) \left(i_2-1+j-r_2\right){\vec{\a}}^{(r_1,r_2)}_{i_1+1,i_2-1,j}\nn
&\hspace{15pt}+(i_1+1)(i_2+1)  s_{11}{\vec{\a}}^{(r_1,r_2)}_{i_1+1,i_2+1,j}\Big]s_{01}^{i_1}s_{0'1}^{i_2}s_{00'}^js_{00}^{r_1-1-i_1-j\over 2}s_{0'0'}^{r_2-1-i_2-j\over 2}{\bf J}\co
\end{align}
and the RHS is
\begin{align}
&\sum_{i_1,i_2,j}\bigg [2r_2\Big((i_1+1) s_{11}{\vec{\a}}^{(r_1,r_2-1)}_{i_1+1,i_2,j}+(j+1) {\vec{\a}}^{(r_1,r_2-1)}_{i_1,i_2-1,j+1} - \left(i_1-1+j-r_1\right){\vec{\a}}^{(r_1,r_2-1)}_{i_1-1,i_2,j}\Big)\nn
&\hspace{15pt}+2r_1\Big((i_2+1) s_{11}{\vec{\a}}^{(r_1-1,r_2)}_{i_1,i_2+1,j}+(j+1) {\vec{\a}}^{(r_1-1,r_2)}_{i_1-1,i_2,j+1}- \left(i_2-1+j-r_2\right){\vec{\a}}^{(r_1-1,r_2)}_{i_1,i_2-1,j}\Big)\nn
&\hspace{15pt}+2r_1r_2\left({\vec{\a}}^{(r_1-1,r_2-1)}_{i_1,i_2,j;\what{3}-\what{1}-\what{2}}
-f_{12}{\vec{\a}}^{(r_1-1,r_2-1)}_{i_1,i_2,j}\right)\bigg]s_{01}^{i_1}s_{0'1}^{i_2}s_{00'}^js_{00}^{r_1-1-i_1-j\over 2}s_{0'0'}^{r_2-1-i_2-j\over 2}{\bf J}\ed
\end{align}
Comparing both sides of equation \eref{2l1l2-recur}, we have
\begin{align}
&(j+1) \left(D+r_1+r_2-j-2\right){\vec{\a}}^{(r_1,r_2)}_{i_1,i_2,j+1}-(i_2+1)  \left(i_1-1+j-r_1\right){\vec{\a}}^{(r_1,r_2)}_{i_1-1,i_2+1,j}\nn
&\hspace{0pt}+\left(i_1+j-1-r_1\right) \left(i_2+j-1-r_2\right) {\vec{\a}}^{(r_1,r_2)}_{i_1,i_2,j-1}-(i_1+1) \left(i_2-1+j-r_2\right){\vec{\a}}^{(r_1,r_2)}_{i_1+1,i_2-1,j}\nn
&\hspace{00pt}+(i_1+1)(i_2+1)  s_{11}{\vec{\a}}^{(r_1,r_2)}_{i_1+1,i_2+1,j}\nn
=~&2r_2\Big((i_1+1) s_{11}{\vec{\a}}^{(r_1,r_2-1)}_{i_1+1,i_2,j}+(j+1) {\vec{\a}}^{(r_1,r_2-1)}_{i_1,i_2-1,j+1}- \left(i_1-1+j-r_1\right){\vec{\a}}^{(r_1,r_2-1)}_{i_1-1,i_2,j}\Big)\nn&+2r_1\Big((i_2+1) s_{11}{\vec{\a}}^{(r_1-1,r_2)}_{i_1,i_2+1,j}+(j+1) {\vec{\a}}^{(r_1-1,r_2)}_{i_1-1,i_2,j+1}- \left(i_2-1+j-r_2\right){\vec{\a}}^{(r_1-1,r_2)}_{i_1,i_2-1,j}\Big)\nn&+2r_1r_2\left({\vec{\a}}^{(r_1-1,r_2-1)}_{i_1,i_2,j;\what{3}-\what{1}-\what{2}}
-f_{12}{\vec{\a}}^{(r_1-1,r_2-1)}_{i_1,i_2,j}\right)\ed \label{dR1dR2-Relation}
\end{align}

By now, we have obtained three $\cal T$-type relations \eref{dR1dR1-Relation}, \eref{dR2dR2-Relation} and \eref{dR1dR2-Relation}. As discussed in the introduction, these relations  are not sufficient to solve all expansion coefficients iteratively. In Table \underline{\ref{number-a-Teqs}}, we list the number of expansion coefficients (each vector $\vec{\a}$ is counted as one
coefficient) and the independent equations given by three ${\cal T}$-type recursions. Comparing two tables, we find the number of reminding unknown terms is universal: $N_{\a}-N_{\cal T}=1$. So for each rank level $r\equiv (r_1+r_2)$ there reminds $(r+1)$ unknown expansion coefficients. We need to find more relations to determine them.
\begin{table}[!htbp]
	\centering
	\renewcommand{\arraystretch}{1}
	\begin{tabular}{|>{\centering\arraybackslash}p{1.23cm}|>{\centering\arraybackslash}p{0.5cm}|>{\centering\arraybackslash}p{0.5cm}|>{\centering\arraybackslash}p{0.5cm}|>{\centering\arraybackslash}p{0.5cm}|>{\centering\arraybackslash}p{0.5cm}|>{\centering\arraybackslash}p{0.5cm}|}
		\hline
		\diagbox{$r_2$}{$N_\a$}{$r_1$} & 0 & 1 & 2 &3&4&5\\
		\hline
		0 & 1 & 1 & 2 & 2 & 3&3\\
		\hline
		1 & 1 & 2 & 3 & 4 & 5&6\\
		\hline
		2 & 2 & 3 & 6 & 7 & 10&11\\%
		\hline
		3&2 & 4 & 7 & 10 & 13&16\\
		\hline
		4&3 & 5 & 10 & 13 & 19&22\\
		\hline
		5&3 & 6 & 11 & 16 & 22 & 28\\
		\hline
	\end{tabular}
    ~
    \renewcommand{\arraystretch}{1.014}
	\begin{tabular}{|>{\centering\arraybackslash}p{1.55cm}|>{\centering\arraybackslash}p{0.5cm}|>{\centering\arraybackslash}p{0.5cm}|>{\centering\arraybackslash}p{0.5cm}|>{\centering\arraybackslash}p{0.5cm}|>{\centering\arraybackslash}p{0.5cm}|>{\centering\arraybackslash}p{0.5cm}|}
		\hline
		\diagbox{$r_2$}{$\ N_{\cal T}\ $}{$r_1$} & 0 & 1 & 2 &3&4&5\\
		\hline
		0 & 0 & 0 & 1 & 1 & 2&2\\
		\hline
		1 & 0 & 1 & 2 & 3 & 4 & 5\\
		\hline
		2 & 1 & 2 & 5 & 6 & 9 & 10\\   %
		\hline
		3&1 & 3 & 6 & 9 & 12 & 15\\
		\hline
		4&2 & 4 & 9 & 12 & 18 & 21 \\
		\hline
		5&2 & 5 & 10 & 15 & 21 & 27\\
		\hline
	\end{tabular}
	\caption{Number of expansion coefficients (left) and independent $\cal T$-type equations (right).}
	\label{number-a-Teqs}
\end{table}
%
%
\subsection{Recursions relation from ${\cal D}$-type operators}

Having used the ${\cal T}$-type operators, from the experience of one-loop integrals, it is obvious that we need to consider the ${\cal D}$-type operators. Naively acting ${\cal D}_{10}$ and ${\cal D}_{10'}$
defined in \eref{D-def} globally to \eref{set-1-2}, we get $K\cdot \ell_i$ in numerator, which are {\sl irreducible scalar products} and
can not be reduced further algebraically. How to get out of the deadlock?
Let us look back to the two-loop integrals. Instead of doing the loop
integration together, one can think it as two
times integration: first, we integrate the $\ell_1$, then we
carry out the integration of $\ell_2$. Using this idea, we can
first do the tensor reduction of  sub one-loop integrals. Let us see what we can get.

Let us start from the sub one-loop integrals. From Appendix \ref{apdix:tadbub},  the $r$-rank tensor reduction of one-loop bubbles
\bea
I_2^{(r)}=\int d\ell {(2 \ell \cdot R)^r\over P_0P_1}
\eea
can be written as summation of $(r-1)$-rank and $(r-2)$-rank bubbles as well as  the contribution of tadpoles, where $P_0\equiv\ell^2-M_0^2,P_1\equiv (\ell-K)^2-M_1^2$.
The striking point is that the Gram determinant $K^2$ only appears once in the denominator, i.e.,
\begin{align}
I_2^{(r)}&={1\over (D+r-3)s_{11}}\bigg[{(D+2(r-2))f_1s_{01}}I_2^{(r-1)}\nn
&\newline -(r-1)\big(4M_0^2s_{01}^2+(f_1^2-4M_0^2s_{11})s_{00}\big )I_2^{(r-2)}+{\cal R}_{Tad}^{(r)}\bigg]~~~\label{bubble-rel}
\end{align}
where we have defined
\bea
s_{11}\equiv K^2\co~~f_1\equiv s_{11}+M_0^2-M_1^2\ed
\eea
For example
\bea
I_{2}^{(2)}
& = & {D f_1 s_{01}\over (D-1) s_{11}}I_2^{(1)}- { 4 M_0^2 s_{01}^2 +s_{00} (f_1^2-4 M_0^2s_{11})\over (D-1)s_{11}}I_2\nn
&&+ {s_{00} f_1\over (D-1) s_{11}}I_{2;\what{1}}+  { 2(D-2) s_{01}^2 +s_{00} (s_{11}-M_0^2+M_1^2)\over (D-1)s_{11}}I_{2;\what{0}}
\eea
where the pole of $s_{11}$ only appears in the overall factor.
Now we insert \eref{bubble-rel} to the two loop integration by regarding propagator $D_1,D_3$ as the propagators $P_0,P_1$  respectively, i.e., we  reduce $\ell_1$ first
\begin{align}
&\int d\ell_2 \int d\ell_1 \intg{r_1}{r_2}=\int d\ell_2 \int d\ell_1 \bigg[{(D+2(r_1-2))\wt{f}_1\wt{s}_{01}\intg{r_1-1}{r_2}\over (D+r_1-3)\wt{s}_{11}}\nn
&\hspace{8pt}-(r_1-1){\big(4M_1^2\wt{s}_{01}^2+(\wt{f}_1^2-4M_1^2\wt{s}_{11})s_{00}\big )\intg{r_1-2}{r_2}\over (D+r_1-3)\wt{s}_{11}}+{\Rterm{r_1}{r_2}{1}\over (D+r_1-3)\wt{s}_{11}}\bigg ]~~~\label{3.19}
\end{align}
where we have defined
\begin{align}
\intg{r_1}{r_2}&\equiv {(2R_1\cdot \ell_1)^{r_1}(2R_2\cdot \ell_1)^{r_2}\over D_1D_2D_3}\co~~~\hatintg{r_1}{r_2}{i}\equiv {D_i(2R_1\cdot \ell_1)^{r_1}(2R_2\cdot \ell_1)^{r_2}\over D_1D_2D_3}\,,
~~~\label{cal-I-def}
\end{align}
and
\begin{align}
\wt{s}_{01}=R_1\cdot (K-\ell_2)\co~~~\wt{s}_{11}=(K-\ell_2)^2\co~~~\wt{f}_1=\wt{s}_{11}+M_1^2-M_3^2\ed~~~\label{f1}
\end{align}
The lower-topology term $\Rterm{r_1}{r_2}{1}$ in \eref{3.19}
has the form ${{\cal N}(\ell_2)\over D_i}\times {(2\ell_2\cdot R_2)^{r_2}\over D_2}$
with $i=1,3$, which is counted by $\hatintg{r_1}{r_2}{i}$.
For example, with rank $r=1$, the tadpole part is
\begin{align}
{\cal I}_{Tad}^{(1)}=(2-D)R\cdot K \left[\int {d\ell \over P_0} - \int {d\ell \over P_1}\right]\co
\end{align}
thus in \eref{3.19} we have
\begin{align}
{\cal R}_{\ell_1,Tad}^{(1,r_2)}=(2-D)R\cdot (K-\ell_2)\left[{(2R_2\cdot \ell_2)^{r_2}\over D_1 D_2}-{(2R_2\cdot \ell_2)^{r_2}\over D_2 D_3}\right]\ed
\end{align}
The appearance $\wt{s}_{11}$ in the denominator in \eref{3.19} causes some trouble since it depends on $\ell_2$, while the original sunset integrals
do not have such a denominator. In the paper \cite{Actis:2004bp}, they keep this factor and regard the two-loop integrals with $\wt{s}_{11}$ in the denominator as
a new master integral. However, according to the reduction idea, such
a solution is somewhat surprising since we expect that the
reduction is achieved by removing original propagators. Thus we should
reach the original topology and its sub topologies at last.

Since we  want to keep the original reduction picture, we will try to get rid of
$\wt{s}_{11}$  in \eref{3.19}. The idea is simple:
one can multiply both sides with $(D+r_1-3)\wt{s}_{11}$ before integrating $\ell_2$ and get
\begin{align}
&\intll (D+r_1-3)\wt{s}_{11}\intg{r_1}{r_2}=\intll  \bigg[{(D+2(r_1-2))\wt{f}_1\wt{s}_{01}}\intg{r_1-1}{r_2}\nn&\hspace{79pt}
-(r_1-1)\big(4M_1^2\wt{s}_{01}^2+(\wt{f}_1^2-4M_1^2\wt{s}_{11})s_{00}\big )\intg{r_1-2}{r_2}+\Rterm{r_1}{r_2}{1}\bigg]\ed\label{loop-by-loop}
\end{align}
Writing
\bea \wt{s}_{11}=(K-\ell_2)^2=K^2+\ell_2^2-2K\cdot \ell_2= s_{11}+D_2+M_2^2-2K\cdot \ell_2\co~~~\label{s11}\eea
the LHS of the equation becomes (where we have suppressed the integral sign $\intll$ for simplicity)
\begin{align}
&\newline(D+r_1-3)\big(s_{11}+D_2+M_2^2-2K\cdot \ell_2\big)\intg{r_1}{r_2}\nn
&
=(D+r_1-3)\left[(s_{11}+M_2^2)\intg{r_1}{r_2}+\hatintg{r_1}{r_2}{2}-{\mathcal{D}_{10'}\over r_2+1}\intg{r_1}{r_2+1}\right]~~~\label{3.26}
\end{align}
where in the last term we replace the new tensor structure $2K\cdot \ell_2$ by the action of  differential operator ${\cal D}_{10'}$ on the basic form $\intg{r_1}{r_2+1}$ (see \eref{cal-I-def}). One important
point in \eref{3.26} is that we have the ${\cal D}$-type action on
$\intg{r_1}{r_2+1}$, which is what we are looking for.

Now we consider the RHS of \eref{loop-by-loop}.
There are three terms: $\wt{f}_1\wt{s}_{01}\intg{r_1-1}{r_2},\wt{s}_{01}^2\intg{r_1-2}{r_2},\big(\wt{f}_1^2-4M_1^2\wt{s}_{11}\big )\intg{r_1-2}{r_2}$.
For the first term, using the expression of $\wt{f}_1$ in \eref{f1}
and the rewriting in \eref{s11}, we can write it as
\begin{align}
&\newline\wt{f}_1\wt{s}_{01}\intg{r_1-1}{r_2}\nn
&=\left(s_{01}-R_1\cdot \ell_2\right)\hatintg{r_1-1}{r_2}{2}+f_{12}s_{01}\intg{r_1-1}{r_2}\nn
&\newline -\left(s_{01}(2K\cdot \ell_2)+f_{12}(R_1\cdot \ell_2)\right)\intg{r_1-1}{r_2}+(2K\cdot \ell_2)(R_1\cdot \ell_2)\intg{r_1-1}{r_2}\nn
&=\left(s_{01}-R_1\cdot \ell_2\right)\hatintg{r_1-1}{r_2}{2}+f_{12}s_{01}\intg{r_1-1}{r_2}\nn
&\newline -\left(2s_{01}K\cdot\ell_2+f_{12}R_1\cdot \ell_2\right)\intg{r_1-1}{r_2}+{\mathcal{D}_{10'}\mathcal{D}_{00'}\over 2(r_2+1)(r_2+2)}\intg{r_1-1}{r_2+2}\ed\nn~~~~\label{3.28}
\end{align}
From the last equation in \eref{3.28} one can see that the rank level
of all terms will be less than  $(r_1+r_2+1)$, except the  last term
with rank level $(r_1+r_2+1)$. For example, the term
$f_{12}R_1\cdot \ell_2\intg{r_1-1}{r_2}={f_{12}\over 2(r_2+1)} {\cal D}_{00'}
\intg{r_1-1}{r_2+1}$ has rank $r_1+r_2$. Similar analysis can be done for other two terms
in the RHS of \eref{loop-by-loop}, it is easy to see that they can be written as a combination of terms with  rank level lower than $(r_1+r_2+1)$.

Collecting all terms together, the RHS of \eref{loop-by-loop} is
\begin{small}
\begin{align}
&-(D+2r_1-4)\left(2s_{01}K\cdot\ell_2+f_{12}R_1\cdot \ell_2\right)\intg{r_1-1}{r_2}+{(D+2r_1-4)\mathcal{D}_{10'}\mathcal{D}_{00'}\over 2(r_2+1)(r_2+2)}\intg{r_1-1}{r_2+2}\nn
&+(D+2r_1-4)\bigg[\left(s_{01}-R_1\cdot \ell_2\right)\hatintg{r_1-1}{r_2}{2}+f_{12}s_{01}\intg{r_1-1}{r_2}\bigg]-(r_1-1)\bigg[4M_1^2(s_{01}-R_1\cdot \ell_2)^2\nn
&+\big[(f_{12}+D_2-2K\cdot \ell_2)^2-4M_1^2(K-\ell_2)^2\big]s_{00}\bigg]\intg{r_1-2}{r_2}+\Rterm{r_1}{r_2}{1}\ed
\end{align}
\end{small}
Rearranging  \eref{loop-by-loop}, we arrive
\begin{align}
&\newline{(D+r_1-3)\mathcal{D}_{10'}\over r_2+1}\intg{r_1}{r_2+1}+{(D+2r_1-4)\mathcal{D}_{10'}\mathcal{D}_{00'}\over 2(r_2+1)(r_2+2)}\intg{r_1-1}{r_2+2}\nn
&=(D+r_1-3)\Big[(s_{11}+M_2^2)\intg{r_1}{r_2}+\hatintg{r_1}{r_2}{2}\Big]+(D+2r_1-4)\Big[\left(2s_{01}K\cdot \ell_2\right)\intg{r_1-1}{r_2}\nn
&\newline-\left(s_{01}-R_1\cdot \ell_2\right)\left(\hatintg{r_1-1}{r_2}{2}+f_{12}\intg{r_1-1}{r_2}\right)\Big]+(r_1-1)\Big[4M_1^2(s_{01}-R_1\cdot \ell_2)^2 \nn
&\newline+\big[(f_{12}+D_2-2K\cdot \ell_2)^2-4M_1^2(K-\ell_2)^2\big]s_{00}\Big]\intg{r_1-2}{r_2}-\Rterm{r_1}{r_2}{1}\ed  \label{loop1-reduce}
\end{align}
The recursion relation \eref{loop1-reduce} is the wanted ${\cal D}$-type recursion relation.
The LHS of the equation can be written as summation of expansion coefficients with rank $(r_1,r_2+1), (r_1-1,r_2+2)$ of rank level $(r_1+r_2+1)$ while the RHS can be written as sum of expansion coefficients with  rank level less than $(r_1+r_2+1)$ and lower topologies. Thus, it is a recursion relation for rank level, instead of
the explicit rank configuration $(r_1,r_2)$.
The reason is that under the ${\cal D}$-type action, different rank configurations are mixed as at the LHS of \eref{loop1-reduce}.

By our recursion assumption,  the RHS is considered to be known, and
we can just write it as
\begin{align}
{\cal B}^{(r_1,r_2)}_{\ell_1}{\bf J}=\sum_{i_1,i_2,j}s_{01}^{i_1}s_{0'1}^{i_2}s_{00'}^js_{00}^{r_1-i_1-j\over 2}s_{0'0'}^{r_2-i_2-j\over 2}\vec{\beta}_{\ell_1;i_1,i_2,j}^{(r_1,r_2)}{\bf J}\ed~~\label{3.31}
\end{align}
The  expression of \eref{3.31} for explicit $r_1,r_2$ can be found
in Appendix \ref{apdix:lowerterm}.
Putting the expansion \eref{coeff-exp} into the LHS and after some algebra, finally we arrive the algebraic recursion relation for unknown coefficients $\vec{\a}$
	\begin{small}
		\allowdisplaybreaks
		\begin{align}
		&{(D+r_1-3)\over r_2+1}\bigg[(j+1) \vec{\a} _{i_1-1,i_2,j+1}^{\left(r_1,r_2+1\right)}+\left(-i_2-j+r_2+2\right) \vec{\a} _{i_1,i_2-1,j}^{\left(r_1,r_2+1\right)}+\left(i_2+1\right) s_{11} \vec{\a} _{i_1,i_2+1,j}^{\left(r_1,r_2+1\right)}\bigg]\nn
		&\newline+{(D+2r_1-4)\over 2(r_2+1)(r_2+2)}\bigg[(j+1) (j+2) \left(D-j+r_1+r_2-2\right) \vec{\a} _{i_1-1,i_2,j+2}^{\left(r_1-1,r_2+2\right)}\nn
		&\newline-(j+1) \left(i_2+j-r_2-2\right) \left(D+i_1-j+r_1+r_2-1\right) \vec{\a} _{i_1,i_2-1,j+1}^{\left(r_1-1,r_2+2\right)}\nn
		&\newline+\left(i_2+1\right) (j+1) \left(D+i_1-j+r_1+r_2-1\right)s_{11} \vec{\a} _{i_1,i_2+1,j+1}^{\left(r_1-1,r_2+2\right)}\nn
		&\newline-\left(i_2+1\right) (j+1) \left(i_1+j-r_1\right) \vec{\a} _{i_1-2,i_2+1,j+1}^{\left(r_1-1,r_2+2\right)}\nn
		&\newline+\left(i_2+j+1\right) \left(i_1+j-r_1\right) \left(i_2+j-r_2-2\right) \vec{\a} _{i_1-1,i_2,j}^{\left(r_1-1,r_2+2\right)}\nn
		&\newline-\left(i_1+j-r_1\right) \left(i_2+j-r_2-4\right) \left(i_2+j-r_2-2\right) \vec{\a} _{i_1,i_2-1,j-1}^{\left(r_1-1,r_2+2\right)}\nn
		&\newline+\left(i_1+1\right) \left(i_2+j-r_2-4\right) \left(i_2+j-r_2-2\right) \vec{\a} _{i_1+1,i_2-2,j}^{\left(r_1-1,r_2+2\right)}\nn
		&\newline-\left(i_2+1\right) \left(i_2+2\right)  \left(i_1+j-r_1\right)s_{11} \vec{\a} _{i_1-1,i_2+2,j}^{\left(r_1-1,r_2+2\right)}\nn
		&\newline+\left(i_2+1\right)  \left(i_1+j-r_1\right) \left(i_2+j-r_2-2\right) s_{11}\vec{\a} _{i_1,i_2+1,j-1}^{\left(r_1-1,r_2+2\right)}\nn
		&\newline-\left(i_1+1\right) \left(2 i_2+1\right) \left(i_2+j-r_2-2\right)s_{11}  \vec{\a} _{i_1+1,i_2,j}^{\left(r_1-1,r_2+2\right)}\nn
		&\newline+\left(i_1+1\right) \left(i_2+1\right) \left(i_2+2\right) s_{11}^2 \vec{\a} _{i_1+1,i_2+2,j}^{\left(r_1-1,r_2+2\right)}\bigg]=\vec{\beta}_{\ell_1;i_1,i_2,j}^{(r_1,r_2)}\label{recur-l1}\ed
		\end{align}
	\end{small}
Similarly, if we reduce $\ell_2$ first, we have
\begin{align}
&\newline{(D+r_2-3)\mathcal{D}_{10}\over r_1+1}\intg{r_1+1}{r_2}+{(D+2r_2-4)\mathcal{D}_{10}\mathcal{D}_{0'0}\over 2(r_1+1)(r_1+2)}\intg{r_1+2}{r_2-1}\nn
&=(D+r_2-3)\Big[(s_{11}+M_1^2)\intg{r_1}{r_2}+\hatintg{r_1}{r_2}{1}\Big]+(D+2r_2-4)\Big[\left(2s_{0'1}K\cdot \ell_1\right)\intg{r_1}{r_2-1}\nn
&\newline-\left(s_{0'1}-R_2\cdot \ell_1\right)\left(\hatintg{r_1}{r_2-1}{1}+f_{12}\intg{r_1}{r_2-1}\right)\Big]-(r_2-1)\Big[4M_2^2(s_{0'1}-R_2\cdot \ell_1)^2\nn
&\newline +\big[(f_{12}+D_1-2K\cdot \ell_1)^2-4M_2^2(K-\ell_1)^2\big]s_{0'0'}\Big]\intg{r_1}{r_2-2}-\Rterm{r_1}{r_2}{2}\co \label{loop2-reduce}
\end{align}
which is the dual expression of \eref{loop1-reduce}.
Similarly, the lower-topology term $\Rterm{r_1}{r_2}{2}$ in the equation can be got from ${\cal R}_{Tad}^{(r_2)}$ by regarding $P_0,P_1$ as $D_2,D_3$ respectively and multiplying with ${(2\ell_1\cdot R_1)^{r_1}/D_1}$. Again, after plugging expansion \eref{coeff-exp} into the LHS of \eref{loop2-reduce} and writing the RHS as
\begin{align}
{\cal B}^{(r_1,r_2)}_{\ell_2}{\bf J}=\sum_{i_1,i_2,j}s_{01}^{i_1}s_{0'1}^{i_2}s_{00'}^js_{00}^{r_1-i_1-j\over 2}s_{0'0'}^{r_2-i_2-j\over 2}\vec{\beta}_{\ell_2;i_1,i_2,j}^{(r_1,r_2)}{\bf J}\co
\end{align}
we will get another algebraic recursion equation for expansion coefficients which is dual to \eref{recur-l1}.

By now, we have derived two $\cal D$-type relations \eref{loop1-reduce} and  \eref{loop2-reduce}.
Unlike the $\cal T$-type relations  which give a recursion relation
for a particular rank configuration, ${\cal D}$-type\footnote{We can understand the ${\cal D}$-type relation from another point of view. In some sense, it likes the IBP relation coming from, for example, $q^{\mu}{\d\over \d \ell_1^{\mu}}$. However, since we have used the  final one-loop reduction results, the solving of these IBP relations has been avoid. Thus the method in this paper is somehow more efficient. } will mix them: one $\cal D$-type relation \eref{loop1-reduce} mixes ranks $(r_1,r_2+1)$ and $(r_1-1,r_2+2)$ while the other $\cal D$-type relation \eref{loop2-reduce} mixes ranks $(r_1+1,r_2)$ and $(r_1+2,r_2-1)$.
When using both types,  one must do reduction level by level. As shown in Table \underline{\ref{DTeqnumbers}}, one can solve all expansion coefficients for $r_1+r_2>2$ by combining $\cal T$-type and $\cal D$-type relations while there are still four expansion coefficients to be determined, which indicates we need more relations for $r_1+r_2\le 2$ level. As will be discussed later, this observation gives us the hint of the choice of master integrals.
\begin{table}
	\centering
	\begin{tabular}{>{\centering\arraybackslash}p{1.9cm}|>{\centering}p{1.9cm}>{\centering}p{1.9cm}>{\centering}p{1.9cm}>{\centering\arraybackslash}p{1.9cm}}
		\hline
		$r_1+r_2$ & $N_{\vec{\a}}$ & $N_{\cal T}$ & $N_{{\cal T}\cup{\cal D}}$ &$N_{\vec{\a}}-N_{{\cal T}\cup{\cal D}}$ \\
		\hline
		0 & 1 & 0 & 0 & 1 \\
		1 & 2 & 0 & 0 & 2 \\
		2 & 6 & 3 & 5 & 1 \\
		3 & 10 & 6 & 10 & 0 \\
		4&20 & 15 & 20 & 0 \\
		5&30 & 24 & 30 & 0 \\
		6&50 & 43 & 50 & 0 \\
		\hline
	\end{tabular}
	\caption{Number of expansion coefficients and independent  equations for several rank levels}
	\label{DTeqnumbers}
\end{table}
\subsection{Master integrals choice}
\label{subsec:MIs}
Up to now, we have avoided giving the explicit choice of master basis in
\eref{coeff-exp}, since all recursion relations of ${\cal T}$-type and
${\cal D}$-type are independent of the choice. The only constraint is that they can not contain auxiliary vectors. However, as pointed out
in Table \underline{\ref{DTeqnumbers}}, the reduction for some lower-rank tensor integrals is not completely clear.

Let us consider them one by one. For the rank level zero,
it is the scalar sunset $I_{1,1,1}$ with no auxiliary vectors, so there are no differential operators that can act on it with nonzero result. Then we must regard it as a master integral, i.e., $J_1\equiv I_{1,1,1}$.
Next, we consider  integrals with rank level one, for example, the integration
\bea I_{1,1,1}[\ell_1]=\int {d\ell_{1,2}\ell_1^\mu \over D_1D_2D_3}\ed\eea
According to its tensor structure, we should have
\bea  I_{1,1,1}[\ell_1]= B K^\mu\ed\eea
The logic of PV-reduction method is to solve $B$ by multiplying $K$ at both sides
\begin{align}
\int {d\ell_{1,2}(2\ell_1 \cdot K) \over D_1D_2D_3}=BK^2\ed \label{rank10-basis}
\end{align}
Now a key point appears. Unlike the one-loop case, the contraction $\ell_1\cdot K$ can not be written as the combinations of $D_i$. Thus, we can not
solve $B$ as a function of external momentum $K$, masses, and space-time dimension $D$ only.
This observation is well known, and it is related to the concept of {\bf irreducible scalar product (ISP)}. For a given $L$-loop integrals with
$E+1$ external legs, there are ${L(L+1)\over 2}+LE$ independent scalar products involving at least one loop momentum. If there are $N$ propagators $D_i$,
$N$ scalar products can be written as the linear combination of $D_i$'s, which leads  $N_{ISP}={L(L+1)\over 2}+LE-N$ irreducible scalar products.
For one-loop integrals, we have $L=1$ and $N=E+1$, so there is no ISP, and every tensor integral can be decomposed to the scalar basis.
But for two-loop integrals, there are nontrivial ISPs. Thus there are multiple master integrals for a given topology. For our sunset topology, we have
$L=2,E=1,N=3$ and $N_{ISP}=2$, which are given by $(\ell_1\cdot K)$ and $(\ell_2\cdot K)$. With the above analysis and the irreducibility of the LHS of
\eref{rank10-basis}, it is natural to take the LHS to be a master integral
\bea
J_2\equiv \int {d\ell_{1,2}(2\ell_1 \cdot K) \over D_1D_2D_3}\,.
\eea
Similarly, we get another master integral
\bea
J_3\equiv \int {d\ell_{1,2}(2\ell_2 \cdot K) \over D_1D_2D_3}\,.
\eea
In Table \underline{\ref{DTeqnumbers}}, there are two undetermined
expansion coefficients for rank level ne. After taking $J_2, J_3$ above to be master
integrals, they can be solved easily as shown in the next section.

Then we consider the rank level two  integrals, for example,
\bea
\int {d\ell_1 d \ell_2 (2\ell_1\cdot R_1)(2\ell_2 \cdot R_2)\over D_1D_2D_3}\,.
\eea
According to it tensor structure, we have
\bea
I_{1,1,1}^{(1,1)}={\vec{\a}}^{(1,1)}_{1,1,0}s_{01}s_{0'1}+{\vec{\a}}^{(1,1)}_{0,0,1}s_{00'}\,.
\eea
The only differential operator that can reduce the integral is ${\cal T}_{00'}$, which gives just one equation while there are two unknown expansion coefficients. As pointed out in Table \underline{\ref{DTeqnumbers}}, there is only one undetermined expansion coefficient for rank level two. When combining the information
of ISP, it is natural to take the integral obtained by acting with $\mathcal{D}_{10} \mathcal{D}_{10'}$, i.e.,
\bea
\int J_4\equiv d\ell_{1,2} {(2K\cdot \ell_1)(2K\cdot \ell_2)\over D_1D_2D_3}\,.
\eea
to be another master integral. In fact, another two choices, i.e.,
\bea \int d\ell_{1,2} {(2K\cdot \ell_1)^2\over D_1D_2D_3}~~~\text{or}~~~\int d\ell_{1,2} {(2K\cdot \ell_2)^2\over D_1D_2D_3}\eea
should be equivalently fine.

Having chosen master integrals for sunset topology,
we discuss the master integrals for lower topologies. These lower topologies are obtained by removing one of the sunset's propagators, and the resulted
topology is the product of two one-loop tadpole integrations.
For one-loop integrals, we know the basis is the scalar integrals, thus
it is easy to see that we have the following three master integrals
\bea
J_5\equiv \int  {d\ell_{1,2}\over D_2D_3},~~~J_6\equiv \int  {d\ell_{1,2}\over D_1D_3},~~~~J_7\equiv\int  {d\ell_{1,2}\over D_1D_2}\ed
\eea
One technical point is that the standard form of $J_{5}, J_{6}$ should
be written with the proper momentum shifting. For example,
\bea J_5&=& \int {d\ell_{1,2}\over (\ell_1^2-M_1^2) ((\ell_1+\ell_2-K)^2-M_3^2)  }= \int {d\ell_{1,2} \over (\ell_1^2-M_1^2) (\ell_2^2-M_3^2)  }\ed~~~\label{3-45}\eea

Our discussion on ISP and the choice of basis hints that the
number ${\cal N}$ of master integrals is related to the number $N_{isp}$ of ISP when the masses and momenta are general.
Another point is that when using the FIRE to do the reduction, it will take a different master basis.
In Appendix \ref{apdix:transmatrix} we will present the non-degenerate transformation matrix between these
two basis, thus  our choice of basis is legitimate.

Before ending this subsection, let us discuss briefly the number of master integrals when the kinematics and masses are not general, for example $K^2=0$ or $M_1=M_2$. 
For $K^2=0$, but $M_1, M_2, M_3$ are different, there are only four master integrals $J_1, J_5,J_6, J_7$
and $J_2, J_3, J_4$ will be decomposed to linear combination of them. We can get the
proper reduction coefficients by taking the limit carefully, as discussed in \cite{Feng:2022rfz}. 
If $M_1=M_2\neq M_3$, but $K^2\neq 0$, we will have $J_5=J_6$ and $J_2=J_3$, i.e., there are only five master integrals. If the reduction coefficients are not singular under the limit
$M_1\to M_2$, we just add reduction coefficients together, i.e., $c_2 J_2+c_3 J_3\to (c_2+c_3) J_2$. If the  reduction coefficients are singular under the limit, we need to 
expand $J_3$ by other master integrals with expansion coefficients as the Taylor
series of $(M_1-M_2)$ as presented in \cite{Feng:2022rfz}.    

\section{Examples}
\label{sec:examples}
Having laid out the general reduction frame in the previous sections, we
present examples to demonstrate our algorithm. In this section, we will
give the exact reduction process for the rank levels from one to four. The
full expansion of coefficients is collected in Appendix \ref{apdix:results} and an attached Mathematica file.

\subsection{Rank level one}
Having chose seven master integrals $J_i$,  we start the reduction for tensor sunset with  rank level  one. There are two integrals
\bea
I_{1,1,1}^{(1,0)}=\int d \ell_{1,2}{(2R_1\cdot \ell_1)\over D_1D_2D_3}\co~~~ I_{1,1,1}^{(0,1)}=\int d \ell_{1,2}{(2R_2\cdot \ell_2)\over D_1D_2D_3}\ed
\eea
Due to the total rank is less than 2, both $\cal T$-type and $\cal D$-type recursions \eref{loop1-reduce} and  \eref{loop2-reduce} can not be used. However, we can apply operators $\mathcal{D}_{10}$ and $\mathcal{D}_{10'}$ on the two integrals respectively.  For example,  the action of $\mathcal{D}_{10}$ on the integral $I_{1,1,1}^{(1,0)}$
gives
\bea
\mathcal{D}_{10} I_{1,1,1}^{(1,0)}=\int d\ell_{1,2} {(2K\cdot \ell_1)\over D_1D_2D_3}=J_2=\vec{\delta}_2 {\bf J} \label{rank10-reduce}
\eea
where we have defined a constant vector
\bea
\vec{\delta}_2=\{0,1,0,0,0,0,0\}\ed
\eea
The reason one can obtain the relation is that we choose $J_2$ as a master integral.

For this case, the expansion \eref{coeff-exp} is
\bea
I_{1,1,1}^{(1,0)}=\vec{\a}^{(1,0)}_{1,0,0}s_{01}{\bf J}\ed
\eea
Plugging it into the LHS of \eref{rank10-reduce}, we have the relation
\bea
\mathcal{D}_{10} I_{1,1,1}^{(1,0)}=\vec{\a}^{(1,0)}_{1,0,0}s_{11}{\bf J}=\vec{\delta}_2 {\bf J}\ed
\eea
Comparing both sides, we can easily solve
\bea
\vec{\a}^{(1,0)}_{1,0,0}={\vec{\delta}_2\over s_{11}}={1\over s_{11}}\{0,1,0,0,0,0,0\}\ed
\eea
Similarly we have
\bea
\vec{\a}^{(0,1)}_{0,1,0}={\vec{\delta}_3\over s_{11}}={1\over s_{11}}\{0,0,1,0,0,0,0\}\ed
\eea
\subsection{Rank level two}
There are three integrals. By the expansion \eref{coeff-exp}, we have
\begin{align}
{\bf C}^{(2,0)}&=\vec{\a} _{0,0,0}^{(2,0)} s_{00}+\vec{\a} _{2,0,0}^{(2,0)} s_{01}^2\co\nn
{\bf C}^{(1,1)}&=\vec{\a} _{0,0,1}^{(1,1)} s_{00'}+\vec{\a} _{1,1,0}^{(1,1)} s_{01} s_{0'1}\co\nn
{\bf C}^{(0,2)}&=\vec{\a} _{0,0,0}^{(0,2)} s_{0'0'}+\vec{\a} _{0,2,0}^{(0,2)} s_{0'1}^2\ed \label{level2-coeff}
\end{align}

Let us start from the rank $(1,1)$. The only $\cal T$-type recursion relation is given by ${\cal T}_{00'}$. Using \eref{dR1dR2-Relation}, we have
\bea
&&{1\over 2}\Big[D{\vec{\a}}^{(1,1)}_{0,0,1}+ s_{11}{\vec{\a}}^{(1,1)}_{1,1,0}\Big]={\vec{\a}}^{(0,0)}_{0,0,0;\what{3}-\what{1}-\what{2}}-f_{12}{\vec{\a}}^{(0,0)}_{0,0,0}+ s_{11}{\vec{\a}}^{(1,0)}_{1,0,0}+ s_{11}{\vec{\a}}^{(0,1)}_{0,1,0}\ed
~~\label{4.9}
\eea
Then we consider the action of ${\cal D}_{10}{\cal D}_{10'}$
\bea
\mathcal{D}_{10}\mathcal{D}_{10'}I_{1,1,1}^{(1,1)}=\int d\ell_{1,2}{(2\ell_1\cdot K)(2\ell_2\cdot K)\over D_1D_2D_3}=J_4=\vec{\delta}_4 {\bf J}\ed \label{rank11-recur2}
\eea
where $\vec{\delta}_4=\{0,0,0,1,0,0,0\}$. Using the expansion \eref{level2-coeff}, we find the LHS is
\bea
\mathcal{D}_{10}\mathcal{D}_{10'}({\vec{\a}}^{(1,1)}_{0,0,1}s_{0'0}+{\vec{\a}}^{(1,1)}_{1,1,0}s_{01}s_{0'1}){\bf J}=s_{11}({\vec{\a}}^{(1,1)}_{0,0,1}+{\vec{\a}}^{(1,1)}_{1,1,0}s_{11}){\bf J}\ed
\eea
Plugging it into \eref{rank11-recur2}, we have
\bea
{\vec{\a}}^{(1,1)}_{0,0,1}+{\vec{\a}}^{(1,1)}_{1,1,0}s_{11}=\vec{\delta}_4/s_{11}\ed~~~\label{4.12}
\eea
Combining \eref{4.9} and \eref{4.12}
 we can solve out
\bea
{\vec{\a}}^{(1,1)}_{0,0,1}&=&{1\over D-1}\bigg[2\Big[{\vec{\a}}^{(0,0)}_{0,0,0;\what{3}-\what{1}-\what{2}}-f_{12}{\vec{\a}}^{(0,0)}_{0,0,0}+ s_{11}{\vec{\a}}^{(1,0)}_{1,0,0}+ s_{11}{\vec{\a}}^{(0,1)}_{0,1,0}\Big]-\vec{\delta}_4/s_{11}\bigg]\co\nn
{\vec{\a}}^{(1,1)}_{1,1,0}&=&{1\over (D-1)s_{11}}\bigg[D\vec{\delta}_4/s_{11}-2\Big[{\vec{\a}}^{(0,0)}_{0,0,0;\what{3}-\what{1}-\what{2}}-f_{12}{\vec{\a}}^{(0,0)}_{0,0,0}+ s_{11}{\vec{\a}}^{(1,0)}_{1,0,0}+ s_{11}{\vec{\a}}^{(0,1)}_{0,1,0}\Big]\bigg]\ed
\eea
Using the results of lower rank level and lower topologies, we get
\bea
{\vec{\a}}^{(1,1)}_{0,0,1}&=&{1\over D-1}\left\{-2 f_{12},2,2,-\frac{1}{s_{11}},-2,-2,2\right\}\co\nn
{\vec{\a}}^{(1,1)}_{1,1,0}&=&{1\over (D-1)s_{11}}\left\{2 f_{12},-2,-2,\frac{D}{s_{11}},2,2,-2\right\}\ed~~~\label{4.14}
\eea
From above analysis, one can see the reason of taking master integral $J_4$ as discussed in previous section.

Next we consider the rank $(2,0)$. The only ${\cal T}$-type relation is given by ${\cal T}_{00}$.  Using
\eref{dR1dR1-Relation} we have
\begin{align}
2D{\vec{\a}}^{(2,0)}_{0,0,0}+2s_{11}{\vec{\a}}^{(2,0)}_{2,0,0}={\vec{\a}}^{(0,0)}_{0,0,0;\what{1}}+M_1^2{\vec{\a}}^{(0,0)}_{0,0,0}\ed \label{rank20-eq1}
\end{align}
The ${\cal D}$-type relation \eref{loop2-reduce} mix the rank $(2,0)$ and the $(1,1)$ and give
\begin{align}
(D-2) \vec{\a} _{0,0,1}^{(1,1)}+\frac{1}{2} (D-2) \vec{\a} _{0,0,0}^{(2,0)}+(D-2) s_{11} \vec{\a} _{1,1,0}^{(1,1)}+\frac{1}{2} (D-2) s_{11} \vec{\a} _{2,0,0}^{(2,0)}=\lowerterm\ed \label{rank20-eq2}
\end{align}
Combining \eref{rank20-eq2} and \eref{rank20-eq1} and using \eref{4.14}, we can solve
\bea
\vec{\a}_{0,0,0}^{(2,0)}&=&{1\over D-1}\left\{2 \left(f_{12}+2 M_1^2\right),-\frac{f_{12}}{s_{11}}-2,-\frac{2 \left(M_1^2+s_{11}\right)}{s_{11}},\frac{2}{s_{11}},4,2,-2\right\}\co\nn
\vec{\a}_{2,0,0}^{(2,0)}&=&{-1\over (D-1)s_{11}}\left\{2 \left(D f_{12}+2 M_1^2\right),\frac{-D \left(f_{12}+2 s_{11}\right)}{s_{11}},\frac{-2 D \left(M_1^2+s_{11}\right)}{s_{11}},\frac{2 D}{s_{11}},4,2 D,-2 D\right\}\ed\nn~~~\label{4.17}
\eea
By proper replacement,  we have
\bea
\vec{\a}_{0,0,0}^{(0,2)}&=&{1\over D-1}\left\{2 \left(f_{12}+2 M_2^2\right),-\frac{2 \left(M_2^2+s_{11}\right)}{s_{11}},-\frac{f_{12}}{s_{11}}-2,\frac{2}{s_{11}},2,4,-2\right\}\co\nn
\vec{\a}_{0,2,0}^{(0,2)}&=&{-1\over (D-1)s_{11}}\left\{2 \left(D f_{12}+2 M_2^2\right),\frac{-2 D \left(M_2^2+s_{11}\right)}{s_{11}},\frac{-D \left(f_{12}+2 s_{11}\right)}{s_{11}},\frac{2 D}{s_{11}},2 D,4,-2 D\right\}\ed\nn
~~~\label{4.18}
\eea
\subsection{Rank level three}
There are four rank configuration and the expansion \eref{coeff-exp}
is
\bea
{\bf C}^{(3,0)}&=&\vec{\a} _{1,0,0}^{(3,0)} s_{00} s_{01}+\vec{\a} _{3,0,0}^{(3,0)} s_{01}^3\co\nn
{\bf C}^{(2,1)}&=&\vec{\a} _{0,1,0}^{(2,1)} s_{00} s_{0'1}+\vec{\a} _{1,0,1}^{(2,1)} s_{01} s_{0'0}+\vec{\a} _{2,1,0}^{(2,1)} s_{01}^2 s_{0'1}\co\nn
{\bf C}^{(1,2)}&=&\vec{\a} _{0,1,1}^{(1,2)} s_{0'0} s_{0'1}+\vec{\a} _{1,0,0}^{(1,2)} s_{01} s_{0'0'}+\vec{\a} _{1,2,0}^{(1,2)} s_{01} s_{0'1}^2\co\nn
{\bf C}^{(0,3)}&=&\vec{\a} _{0,1,0}^{(0,3)} s_{0'1} s_{0'0'}+\vec{\a} _{0,3,0}^{(0,3)} s_{0'1}^3\ed
\eea
We list 14 linear equations of $\cal T$-type and $\cal D$-type relations for these ten expansion coefficients. There are only ten of them are
linear independent and other four extra can be used as the self-consistence check.
We list $\cal T$-type equations first,
\begin{itemize}
	\item $\cal T$-type relations
	\begin{itemize}
		\item rank (3,0)
		\begin{align}
	{\cal T}_{00}:(D+2) \vec{\a} _{1,0,0}^{(3,0)}+3 s_{11} \vec{\a} _{3,0,0}^{(3,0)}=\lowerterm\ed
		\end{align}
		\item rank (2,1)
		\begin{align}
		 {\cal T}_{00}&:D \vec{\a} _{0,1,0}^{(2,1)}+\vec{\a} _{1,0,1}^{(2,1)}+s_{11} \vec{\a} _{2,1,0}^{(2,1)}=\lowerterm\co\nn
		 {\cal T}_{00'}&:(D+1) \vec{\a} _{1,0,1}^{(2,1)}+2 \vec{\a} _{0,1,0}^{(2,1)}+2 s_{11} \vec{\a} _{2,1,0}^{(2,1)}=\lowerterm\ed
		\end{align}
		\item rank (1,2)
		\begin{align}
		{\cal T}_{00}&:2 s_{01} \left(D \vec{\a} _{1,0,0}^{(1,2)}+\vec{\a} _{0,1,1}^{(1,2)}+s_{11} \vec{\a} _{1,2,0}^{(1,2)}\right)=\lowerterm\co\nn
		{\cal T}_{00'}&: (D+1) \vec{\a} _{0,1,1}^{(1,2)}+2 \vec{\a} _{1,0,0}^{(1,2)}+2s_{11} \vec{\a} _{1,2,0}^{(1,2)}=\lowerterm\ed
		\end{align}
		\item rank (0,3)
		\begin{align}
		{\cal T}_{0'0'}&:(D+2) \vec{\a} _{0,1,0}^{(0,3)}+3 s_{11} \vec{\a} _{0,3,0}^{(0,3)}=\lowerterm\ed
		\end{align}
	\end{itemize}
\end{itemize}
These six independent equations restrict the number of unknown expansion coefficients to 4. For every rank configuration, the number of $\cal T$  equations is always one less than the number of expansion coefficients, which agrees with our discussion.

Now we consider the $\cal D$ relations for rank level three, which are represented by four arcs in Figure \underline{\ref{D-relation-Figure}}, where two red arcs for reducing $\ell_1$ first and other two orange arcs for reducing $\ell_2$ first.
We list them below:
\begin{itemize}
	\item ${\cal D}$-type relations by reducing $\ell_1$ first
	\begin{itemize}
		\item Mix rank $(0,3)$ and $(1,2)$
	\begin{align}
	&\vec{\a} _{0,1,0}^{(0,3)}+3 \vec{\a} _{0,1,1}^{(1,2)}=\lowerterm\co\nn
	&2 \vec{\a} _{0,1,0}^{(0,3)}+3\vec{\a} _{0,1,1}^{(1,2)}+6 \vec{\a} _{1,0,0}^{(1,2)}+3s_{11} \vec{\a} _{0,3,0}^{(0,3)}+6 s_{11} \vec{\a} _{1,2,0}^{(1,2)}=\lowerterm\ed
	\end{align}
	The reason of having two equations is that even with the removing of one $R_2$
	by ${\cal D}$ operator, we are still left two different tensor structures, i.e., $s_{0'0'}$ and $s_{0'1}^2$.
	
	\item Mix rank $(1,2)$ and $(2,1)$
	\begin{align}
	&\frac{1}{4} D s_{11} \vec{\a} _{0,1,1}^{(1,2)}+(D-1) s_{11} \vec{\a} _{0,1,0}^{(2,1)}=\lowerterm\co\nn
	&\frac{1}{4} D \vec{\a} _{0,1,1}^{(1,2)}+\frac{1}{2} D \left(\vec{\a} _{1,0,0}^{(1,2)}+s_{11} \vec{\a} _{1,2,0}^{(1,2)}\right)+(D-1) \left(\vec{\a} _{1,0,1}^{(2,1)}+s_{11} \vec{\a} _{2,1,0}^{(2,1)}\right)=\lowerterm\ed
	\end{align}
   \end{itemize}
   \item ${\cal D}$-type relations by reducing $\ell_2$ first
   \begin{itemize}
   	\item Mix rank $(3,0)$ and $(2,1)$
   	\begin{align}
   	&3 \vec{\a} _{1,0,1}^{(2,1)}+\vec{\a} _{1,0,0}^{(3,0)}=\lowerterm\co\nn
   	&6 \vec{\a} _{0,1,0}^{(2,1)}+3 \vec{\a} _{1,0,1}^{(2,1)}+2 \vec{\a} _{1,0,0}^{(3,0)}+6 s_{11} \vec{\a} _{2,1,0}^{(2,1)}+3 s_{11} \vec{\a} _{3,0,0}^{(3,0)}=\lowerterm\ed
   	\end{align}
   	\item Mix rank $(1,2)$ and $(2,1)$
   	\begin{align}
   	&(D-1) s_{11} \vec{\a} _{1,0,0}^{(1,2)} +\frac{1}{4} D s_{11} \vec{\a} _{1,0,1}^{(2,1)}=\lowerterm\co \nn
   	&(D-1) \left(\vec{\a} _{0,1,1}^{(1,2)}+s_{11} \vec{\a} _{1,2,0}^{(1,2)}\right)+\frac{D}{4}  \left(2 \vec{\a} _{0,1,0}^{(2,1)}+\vec{\a} _{1,0,1}^{(2,1)}+2 s_{11} \vec{\a} _{2,1,0}^{(2,1)}\right)=\lowerterm\ed
   	\end{align}
   \end{itemize}
\end{itemize}
One can pick ten independent equations from the relations above and solve all expansion coefficients. Denoting
\bea
\vec{x}^{3-\text{level}}&=&\left\{\vec{\a} _{1,0,0}^{(3,0)},\vec{\a} _{3,0,0}^{(3,0)},\vec{\a} _{0,1,0}^{(2,1)},\vec{\a} _{1,0,1}^{(2,1)},\vec{\a} _{2,1,0}^{(2,1)}
,\vec{\a} _{0,1,0}^{(0,3)},\vec{\a} _{0,3,0}^{(0,3)},\vec{\a} _{1,0,0}^{(1,2)},\vec{\a} _{0,1,1}^{(1,2)},\vec{\a} _{1,2,0}^{(1,2)}\right\}\co
\eea
we have linear equations $W\vec{x}^{3-\text{level}}=b$ with matrix $W$
\begin{align}
\left(
\begin{array}{cccccccccc}
	1 & 0 & 0 & 3 & 0 & 0 & 0 & 0 & 0 & 0 \\
	\frac{1}{3} & \frac{s_{11}}{2} & 1 & \frac{1}{2} & s_{11} & 0 & 0 & 0 & 0 & 0 \\
	0 & 0 & -\frac{D}{4} & -\frac{1}{4} & -\frac{s_{11}}{4} & 0 & 0 & 0 & 0 & 0 \\
	0 & 0 & 0 & 0 & 0 & 1 & 0 & 0 & 3 & 0 \\
	0 & 0 & 0 & 0 & 0 & \frac{1}{3} & \frac{s_{11}}{2} & 1 & \frac{1}{2} & s_{11} \\
	0 & 0 & 0 & 0 & 0 & 0 & 0 & -\frac{D}{4} & -\frac{1}{4} & -\frac{s_{11}}{4} \\
	0 & 0 & 1-D & 0 & 0 & 0 & 0 & 0 & -\frac{D }{4} & 0 \\
	0 & 0 & 0 & 1-D & s_{11}-D s_{11} & 0 & 0 & -\frac{D}{2} & -\frac{D}{4} & -\frac{D s_{11}}{2} \\
	0 & 0 & 0 & -\frac{D}{4} & 0 & 0 & 0 & 1-D  & 0 & 0 \\
	0 & 0 & -\frac{D}{2} & -\frac{D}{4} & -\frac{D s_{11}}{2} & 0 & 0 & 0 & 1-D & s_{11}-D s_{11} \\
\end{array}
\right)\co
\end{align}
and the known vector $b$  which corresponds to lower rank level and lower topologies.  We solve these linear equations and present final results in Appendix \ref{apdix:results}.

\subsection{Rank level four}
There are five rank configuration and the expansion \eref{coeff-exp} is
\bea
{\bf C}^{(4,0)}&=&\vec{\a} _{0,0,0}^{(4,0)} s_{00}^2+\vec{\a} _{2,0,0}^{(4,0)} s_{00} s_{01}^2+\vec{\a} _{4,0,0}^{(4,0)} s_{01}^4\co\nn
{\bf C}^{(3,1)}_{1,1,1}&=&\vec{\a} _{0,0,1}^{(3,1)} s_{00} s_{00'}+\vec{\a} _{1,1,0}^{(3,1)} s_{00} s_{01} s_{0'1}+\vec{\a} _{2,0,1}^{(3,1)} s_{01}^2 s_{00'}+\vec{\a} _{3,1,0}^{(3,1)} s_{01}^3 s_{0'1}\nn
{\bf C}^{(2,2)}&=&\vec{\a} _{0,0,2}^{(2,2)} s_{00'}^2+\vec{\a} _{0,2,0}^{(2,2)} s_{00} s_{0'1}^2+\vec{\a} _{1,1,1}^{(2,2)} s_{01} s_{00'} s_{0'1}+\vec{\a} _{2,2,0}^{(2,2)} s_{01}^2 s_{0'1}^2+\vec{\a} _{0,0,0}^{(2,2)} s_{00} s_{0'0'}+\vec{\a} _{2,0,0}^{(2,2)} s_{01}^2 s_{0'0'}\co\nn
{\bf C}^{(1,3)}&=&\vec{\a} _{0,0,1}^{(1,3)} s_{00'} s_{0'0'}+\vec{\a} _{0,2,1}^{(1,3)} s_{00'} s_{0'1}^2+\vec{\a} _{1,1,0}^{(1,3)} s_{01} s_{0'1} s_{0'0'}+\vec{\a} _{1,3,0}^{(1,3)} s_{01} s_{0'1}^3\co\nn
{\bf C}^{(0,4)}&=&\vec{\a} _{0,0,0}^{(0,4)} s_{0'0'}^2+\vec{\a} _{0,2,0}^{(0,4)} s_{0'1}^2 s_{0'0'}+\vec{\a} _{0,4,0}^{(0,4)} s_{0'1}^4\ed
\eea
One can get 34 linear equations from $\cal T$-type and $\cal D$-type relations for these 20 expansion coefficients. We can pick 20 independent equations among them.
There are 18 $\cal T$-type equations
\begin{subequations}
	\allowdisplaybreaks
	\begin{small}
			\allowdisplaybreaks
		\begin{align}
		\bigg \{&4 D \vec{\a} _{0,0,0}^{(0,4)}+8 \vec{\a} _{0,0,0}^{(0,4)}+2 s_{11} \vec{\a} _{0,2,0}^{(0,4)},2 D \vec{\a} _{0,2,0}^{(0,4)}+8 \vec{\a} _{0,2,0}^{(0,4)}+12 s_{11} \vec{\a} _{0,4,0}^{(0,4)},2 D \vec{\a} _{0,0,1}^{(1,3)}+4 \vec{\a} _{0,0,1}^{(1,3)}+2 s_{11} \vec{\a} _{0,2,1}^{(1,3)},\nn
		& 2 D \vec{\a} _{1,1,0}^{(1,3)}+4 \vec{\a} _{0,2,1}^{(1,3)}+4 \vec{\a} _{1,1,0}^{(1,3)}+6 s_{11} \vec{\a} _{1,3,0}^{(1,3)},D \vec{\a} _{0,0,1}^{(1,3)}+2 \vec{\a} _{0,0,1}^{(1,3)}+s_{11} \vec{\a} _{1,1,0}^{(1,3)},\nn
		& D \vec{\a} _{0,2,1}^{(1,3)}+2 \vec{\a} _{0,2,1}^{(1,3)}+2 \vec{\a} _{1,1,0}^{(1,3)}+3 s_{11} \vec{\a} _{1,3,0}^{(1,3)},2 D \vec{\a} _{0,0,0}^{(2,2)}+2 \vec{\a} _{0,0,2}^{(2,2)}+2 s_{11} \vec{\a} _{2,0,0}^{(2,2)},\nn
		& 2 D \vec{\a} _{0,2,0}^{(2,2)}+2 \vec{\a} _{1,1,1}^{(2,2)}+2 s_{11} \vec{\a} _{2,2,0}^{(2,2)},2 D \vec{\a} _{0,0,0}^{(2,2)}+2 \vec{\a} _{0,0,2}^{(2,2)}+2 s_{11} \vec{\a} _{0,2,0}^{(2,2)},2 D \vec{\a} _{2,0,0}^{(2,2)}+2 \vec{\a} _{1,1,1}^{(2,2)}+2 s_{11} \vec{\a} _{2,2,0}^{(2,2)},\nn
		& 2 D \vec{\a} _{0,0,2}^{(2,2)}+4 \vec{\a} _{0,0,0}^{(2,2)}+2 \vec{\a} _{0,0,2}^{(2,2)}+s_{11} \vec{\a} _{1,1,1}^{(2,2)},D \vec{\a} _{1,1,1}^{(2,2)}+4 \vec{\a} _{0,2,0}^{(2,2)}+2 \vec{\a} _{1,1,1}^{(2,2)}+4 \vec{\a} _{2,0,0}^{(2,2)}+4 s_{11} \vec{\a} _{2,2,0}^{(2,2)},\nn
		& 2 D \vec{\a} _{0,0,1}^{(3,1)}+4 \vec{\a} _{0,0,1}^{(3,1)}+2 s_{11} \vec{\a} _{2,0,1}^{(3,1)},2 D \vec{\a} _{1,1,0}^{(3,1)}+4 \vec{\a} _{1,1,0}^{(3,1)}+4 \vec{\a} _{2,0,1}^{(3,1)}+6 s_{11} \vec{\a} _{3,1,0}^{(3,1)},\nn
		& D \vec{\a} _{0,0,1}^{(3,1)}+2 \vec{\a} _{0,0,1}^{(3,1)}+s_{11} \vec{\a} _{1,1,0}^{(3,1)},D \vec{\a} _{2,0,1}^{(3,1)}+2 \vec{\a} _{1,1,0}^{(3,1)}+2 \vec{\a} _{2,0,1}^{(3,1)}+3 s_{11} \vec{\a} _{3,1,0}^{(3,1)},4 D \vec{\a} _{0,0,0}^{(4,0)}+8 \vec{\a} _{0,0,0}^{(4,0)}+2 s_{11} \vec{\a} _{2,0,0}^{(4,0)},\nn
		& 2 D \vec{\a} _{2,0,0}^{(4,0)}+8 \vec{\a} _{2,0,0}^{(4,0)}+12 s_{11} \vec{\a} _{4,0,0}^{(4,0)}\bigg\}=\lowerterm\ed
		\end{align}
	\end{small}
\end{subequations}
For 16 $\cal D$-type equations, we have
\begin{itemize}
	\item reducing $\ell_1$ first
\begin{small}
		\allowdisplaybreaks
\begin{align}
\bigg\{&\frac{1}{3} (D-2) \vec{\a} _{0,0,0}^{(0,4)}+\frac{2}{3} (D-2) \vec{\a} _{0,0,1}^{(1,3)}+\frac{1}{6} (D-2) s_{11} \vec{\a} _{0,2,0}^{(0,4)}+\frac{2}{3} (D-2) s_{11} \vec{\a} _{0,2,1}^{(1,3)},\nn
&\frac{1}{6} (D-2) \vec{\a} _{0,0,0}^{(0,4)}+\frac{1}{3} (D-2) \vec{\a} _{0,0,1}^{(1,3)}+\frac{1}{12} (D-2) s_{11} \vec{\a} _{0,2,0}^{(0,4)}+\frac{1}{3} (D-2) s_{11} \vec{\a} _{1,1,0}^{(1,3)},\nn
&\frac{1}{4} (D-2) \vec{\a} _{0,2,0}^{(0,4)}+\frac{1}{3} (D-2) \vec{\a} _{0,2,1}^{(1,3)}+\frac{2}{3} (D-2) \vec{\a} _{1,1,0}^{(1,3)}+\frac{1}{2} (D-2) s_{11} \vec{\a} _{0,4,0}^{(0,4)}+(D-2) s_{11} \vec{\a} _{1,3,0}^{(1,3)},\nn
&\frac{1}{6} D \vec{\a} _{0,0,1}^{(1,3)}+(D-1) \vec{\a} _{0,0,0}^{(2,2)}+\frac{1}{6} D s_{11} \vec{\a} _{0,2,1}^{(1,3)}+(D-1) s_{11} \vec{\a} _{0,2,0}^{(2,2)},\nn
&\frac{1}{3} D \vec{\a} _{0,0,1}^{(1,3)}+(D-1) \vec{\a} _{0,0,2}^{(2,2)}+\frac{1}{6} D s_{11} \vec{\a} _{0,2,1}^{(1,3)}+\frac{1}{6} D s_{11} \vec{\a} _{1,1,0}^{(1,3)}+\frac{1}{2} (D-1) s_{11} \vec{\a} _{1,1,1}^{(2,2)},\nn
&\frac{1}{6} D \vec{\a} _{0,2,1}^{(1,3)}+\frac{1}{3} D \vec{\a} _{1,1,0}^{(1,3)}+\frac{1}{2} (D-1) \vec{\a} _{1,1,1}^{(2,2)}+(D-1) \vec{\a} _{2,0,0}^{(2,2)}+\frac{1}{2} D s_{11} \vec{\a} _{1,3,0}^{(1,3)}+(D-1) s_{11} \vec{\a} _{2,2,0}^{(2,2)},\nn
&\frac{1}{2} (D+2) \vec{\a} _{0,0,0}^{(2,2)}+\frac{1}{2} (D+2) \vec{\a} _{0,0,2}^{(2,2)}+D \vec{\a} _{0,0,1}^{(3,1)}+\frac{1}{2} (D+2) s_{11} \vec{\a} _{0,2,0}^{(2,2)}+\frac{1}{4} (D+2) s_{11} \vec{\a} _{1,1,1}^{(2,2)}+D s_{11} \vec{\a} _{1,1,0}^{(3,1)},\nn
&\frac{1}{4} (D+2) \vec{\a} _{1,1,1}^{(2,2)}+\frac{1}{2} (D+2) \vec{\a} _{2,0,0}^{(2,2)}+D \vec{\a} _{2,0,1}^{(3,1)}+\frac{1}{2} (D+2) s_{11} \vec{\a} _{2,2,0}^{(2,2)}+D s_{11} \vec{\a} _{3,1,0}^{(3,1)}\bigg\}=\lowerterm\ed
\end{align}
\end{small}
  \item reducing $\ell_2$ first

  	\begin{small}
  			\allowdisplaybreaks
  		\begin{align}
  		\allowdisplaybreaks
  		\bigg\{&D \vec{\a} _{0,0,1}^{(1,3)}+\frac{1}{2} (D+2) \vec{\a} _{0,0,0}^{(2,2)}+\frac{1}{2} (D+2) \vec{\a} _{0,0,2}^{(2,2)}+D s_{11} \vec{\a} _{1,1,0}^{(1,3)}+\frac{1}{4} (D+2) s_{11} \vec{\a} _{1,1,1}^{(2,2)}+\frac{1}{2} (D+2) s_{11} \vec{\a} _{2,0,0}^{(2,2)},\nn&D \vec{\a} _{0,2,1}^{(1,3)}+\frac{1}{2} (D+2) \vec{\a} _{0,2,0}^{(2,2)}+\frac{1}{4} (D+2) \vec{\a} _{1,1,1}^{(2,2)}+D s_{11} \vec{\a} _{1,3,0}^{(1,3)}+\frac{1}{2} (D+2) s_{11} \vec{\a} _{2,2,0}^{(2,2)},\nn&(D-1) \vec{\a} _{0,0,2}^{(2,2)}+\frac{1}{3} D \vec{\a} _{0,0,1}^{(3,1)}+\frac{1}{2} (D-1) s_{11} \vec{\a} _{1,1,1}^{(2,2)}+\frac{1}{6} D s_{11} \vec{\a} _{1,1,0}^{(3,1)}+\frac{1}{6} D s_{11} \vec{\a} _{2,0,1}^{(3,1)},\nn&(D-1) \vec{\a} _{0,0,0}^{(2,2)}+\frac{1}{6} D \vec{\a} _{0,0,1}^{(3,1)}+(D-1) s_{11} \vec{\a} _{2,0,0}^{(2,2)}+\frac{1}{6} D s_{11} \vec{\a} _{2,0,1}^{(3,1)},\nn&(D-1) \vec{\a} _{0,2,0}^{(2,2)}+\frac{1}{2} (D-1) \vec{\a} _{1,1,1}^{(2,2)}+\frac{1}{3} D \vec{\a} _{1,1,0}^{(3,1)}+\frac{1}{6} D \vec{\a} _{2,0,1}^{(3,1)}+(D-1) s_{11} \vec{\a} _{2,2,0}^{(2,2)}+\frac{1}{2} D s_{11} \vec{\a} _{3,1,0}^{(3,1)},\nn&\frac{1}{3} (D-2) \vec{\a} _{0,0,1}^{(3,1)}+\frac{1}{6} (D-2) \vec{\a} _{0,0,0}^{(4,0)}+\frac{1}{3} (D-2) s_{11} \vec{\a} _{1,1,0}^{(3,1)}+\frac{1}{12} (D-2) s_{11} \vec{\a} _{2,0,0}^{(4,0)},\nn&\frac{2}{3} (D-2) \vec{\a} _{0,0,1}^{(3,1)}+\frac{1}{3} (D-2) \vec{\a} _{0,0,0}^{(4,0)}+\frac{2}{3} (D-2) s_{11} \vec{\a} _{2,0,1}^{(3,1)}+\frac{1}{6} (D-2) s_{11} \vec{\a} _{2,0,0}^{(4,0)},\nn&\frac{2}{3} (D-2) \vec{\a} _{1,1,0}^{(3,1)}+\frac{1}{3} (D-2) \vec{\a} _{2,0,1}^{(3,1)}+\frac{1}{4} (D-2) \vec{\a} _{2,0,0}^{(4,0)}+(D-2) s_{11} \vec{\a} _{3,1,0}^{(3,1)}+\frac{1}{2} (D-2) s_{11} \vec{\a} _{4,0,0}^{(4,0)}\bigg\}\nn
  		&=\lowerterm\ed
  		\end{align}
  	\end{small}
\end{itemize}
Picking  20 independent equations, we can solve all expansion coefficients of rank level four. Since its expression is long,
we present them in an attached Mathematica file. All coefficients have
been checked with the analytic output of FIRE~\cite{Smirnov:2008iw,Smirnov:2014hma,Smirnov:2019qkx}.

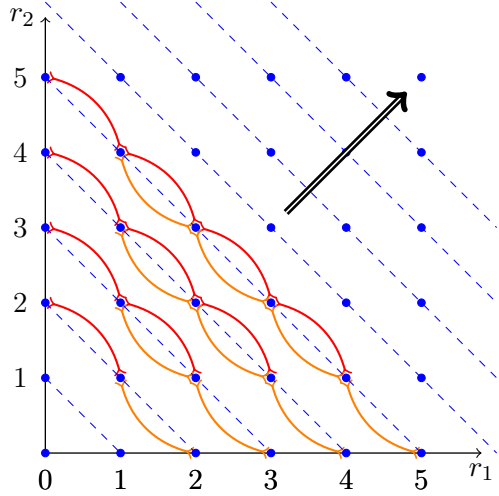
\begin{figure}
	\begin{center}
		\begin{tikzpicture}[scale=1.0]
		\newcommand{\rectangle}[4]{\draw[fill,black] (#1-0.5*#3,#2-0.5*#4)--(#1+0.5*#3,#2-0.5*#4)--(#1+0.5*#3,#2+0.5*#4)--(#1-0.5*#3,#2+0.5*#4)--cycle;}
		\def\size{0.05}
		\def\boxsize{0.07}
		\def\offset{0.04}
		\newcommand{\reduceLeft}[3]{\draw[line width=0.8,#3,>-<](#1,#2+\offset)[bend right]to(#1-1+\offset,#2+1);
		}
		\newcommand{\reduceRight}[3]{\draw[line width=0.8,#3,>-<](#1,#2-\offset)[bend right]to(#1+1-\offset,#2-1);
		}
		\draw[line width=0.5,->](0,0)--(5.8,0);
		\draw[line width=0.5,->](0,0)--(0,5.8);
		\foreach \x in {0,1,2,...,5}{
			\node[below] at (\x,-0.1) {\x};
			\ifnum \x>0
			\node[left] at (-0.1,\x) {\x};
			\fi
			\node[below] at (\x,-0.1) {\x};
			\foreach \y in {0,1,2,...,5}{
				\draw[fill,blue] (\x,\y) circle [radius=\size];
			}

		}
		\node[below] at (5.8,0) {$r_1$};
		\node[left] at (0,5.8) {$r_2$};
		\reduceLeft{1}{1}{red};
		\reduceRight{1}{1}{orange};
		\reduceLeft{1}{2}{red};
		\reduceLeft{2}{1}{red};
		\reduceRight{1}{2}{orange};
		\reduceRight{2}{1}{orange};
		\reduceLeft{1}{3}{red};
		\reduceLeft{2}{2}{red};
		\reduceLeft{3}{1}{red};
		\reduceRight{1}{3}{orange};
		\reduceRight{2}{2}{orange};
		\reduceRight{3}{1}{orange};
		\reduceLeft{1}{4}{red};
		\reduceLeft{2}{3}{red};
		\reduceLeft{3}{2}{red};
		\reduceLeft{4}{1}{red};
		\reduceRight{1}{4}{orange};
		\reduceRight{2}{3}{orange};
		\reduceRight{3}{2}{orange};
		\reduceRight{4}{1}{orange};
		\draw[dashed,blue](0,1)--(1,0);
		\draw[dashed,blue](0,2)--(2,0);
		\draw[dashed,blue](0,3)--(3,0);
		\draw[dashed,blue](0,4)--(4,0);
		\draw[dashed,blue](0,5)--(5,0);
		\draw[dashed,blue](0,6)--(6,0);
		\draw[dashed,blue](1,6)--(6,1);
		\draw[dashed,blue](2,6)--(6,2);
		\draw[dashed,blue](3,6)--(6,3);
		\draw[line width=1,->,double](3.2,3.2)--(4.8,4.8);
		
		\end{tikzpicture}
		
		\caption{Algorithm for tensor reduction of sunset topology, where every point represents a particular tensor configuration $(r_1, r_2)$. After using the $\cal T$-type relations, only one unknown coefficient is left for each point. To determine the last unknown piece, we need to use $\cal D$-type relations to relate different tensor configurations with the same rank level. We have used the red lines to represent relations coming from $\cal D$-type relation for $\ell_1$, and orange lines, for $\ell_2$ respectively (here we draw $\cal D$-type relations for $r_1+r_2\le 5$). Each line provides a nontrivial relation among unknown coefficients. One can see that for rank level $r\geq 3$, the number of lines is equal or larger than the number of unknown coefficients, thus they will give enough equations.
		}
		\label{D-relation-Figure}
	\end{center}
\end{figure}

Before ending this section, let us  summarize the algorithm to reduce tensor integrals of sunset topology to the master basis. The number of expansion coefficients for a particular rank level $r$ is given by
\bea
N_{\vec{a}}[r]=\sum_{r_1+r_2=r}\sum_{j=0}^{\min \{r_1,r_2\}}N_c(r_1-j)N_c(r_2-j)
\eea
where $N_c(m)=\lfloor {m\over 2}\rfloor+1$ is the number of expansion coefficients for a one-loop rank-$m$ tensor bubble. The recursion procedure can be nicely represented in Figure \underline{\ref{D-relation-Figure}}.

\section{Conclusion and outlook}
\label{sec:conclusion}
In this paper, we have taken the first step in  generalizing our
improved PV-reduction method with auxiliary vector $R$ to two-loop
integrals, i.e., we have carried out the tensor reduction for the simplest sunset topology. For two-loop integrals, the $\cal T$-type relations can be established straightforwardly, while the $\cal D$-type relations are rather nontrivial. We need to consider the reduction of its sub one-loop integrals first. Combining the two types of relations, one can solve all expansion coefficients with the proper choice of master basis. 

One of our motivation of the paper is to find an alternative method for higher loop tensor reduction other than IBP method. Comparing these two methods, one can find some
similarities. All IBP relations are generated by following six primary relations
$ \int {\d\over \d \ell_i} \cdot K_j$
with $i=1,2$ and $K_j=(\ell_1, \ell_2, K)$. Similarly, in our algorithm, there are also
six operators $\W K_j\cdot {\d\over \D R_i}$ with $i=1,2$ and $\W K_j=(R_1, R_2, K)$. 
The one to one mapping between these two types of manipulations seems to hint a deep connection between these two methods although we are not clear at this moment. As pointed out in the paper, in our algorithm
we have avoid the appearance of higher power of propagators, thus the number of equations one needs to solve is much fewer than that using standard IBP method.
However, in \cite{Gluza:2010ws} an idea has been suggested to avoid the higher power of propagators in the IBP method. It will be interesting to see if such an idea can be used
in our algorithm. 
 

There are several interesting questions coming from this work.
The first one is that, in section \ref{sec:recursions}, we have discussed the choice of master integrals to be evaluating the ISP. We hope that for arbitrary two-loop or higher loop integrals with general masses and momenta $K$, one can determine the master basis from this perspective.

The second question is the following. As emphasized several times, the nontrivial ${\cal D}$-type relations depend crucially on the reduction relation \eref{bubble-rel} of one-loop integrals. We have tested the reduction of triangles and found similar reduction relation also holds for several examples.
We find such a reduction relation holds for general one-loop integrals with the proof being presented in \cite{Feng:2022rfz}.

The third problem is obvious. In this paper, we focus on
the sunset topology. There are other topologies for two-loop integrals.
One needs to check if the algorithm present in this paper can be generalized to these cases. With results from higher topologies, we can
reduce sunset with propagators having higher power using the idea given in \cite{Feng:2022uqp}. We hope to give a positive answer to these
interesting questions in our future work.

\section*{Acknowledgments}
This work is supported by  Chinese NSF funding under Grant No.11935013, No.11947301, No.12047502 (Peng Huanwu Center).
%
\appendix
\section{The reduction of one-loop tadpoles and bubbles}
\label{apdix:tadbub}
In this part, we will collect the reduction of one-loop tensor tadpoles and bubbles, which has
been extensively used in this paper.

For the one-loop rank $r$ tensor tadpole
\bea I_1^{(r)}\equiv\int {d^D \ell\over (2\pi)^D}  {(2R\cdot \ell)^r\over (\ell^2-M_0^2)},~~~~\label{oneloop-tad-1}
\eea
the reduction result is given by
\bea I_1^{(r)}= C_{1\to 1}^{(r)}\int {d^D \ell\over (2\pi)^D}  {1\over (\ell^2-M_0^2)},~~~~\label{oneloop-tad-2}\eea
where
\bea C^{(r)}_{1\to 1}=\gamma(r,M_0)(R^2)^{r\over 2},~~~~ \gamma(r,M_0)\equiv {1+(-)^r\over 2} {(r-1)!!2^r M_0^{r}\over \prod_{i=1}^{r\over 2} D+2(i-1)}.~~~~\label{oneloop-tad-3}\eea

For the one-loop rank $r$ tensor bubble
\bea I_2^{(r)}\equiv\int {d^D \ell\over (2\pi)^D}  {(2R\cdot \ell)^r\over (\ell^2-M_0^2)((\ell-K)^2-M_1^2)},~~~~\label{oneloop-bub-1}
\eea
the reduction result is given by
\bea I_2^{(r)}& = &C^{(r)}_{2\to 2}I_2+ \sum_{i=0,1}C^{(r)}_{2\to 2;\what{i}} I_{2;\what{i}}~,~~~~\label{oneloop-bub-2}\eea
where
\bea I_2=I_2^{(0)},~~~I_{2;\what{0}}=\int {d^D \ell\over (2\pi)^D}  {1\over (\ell^2-M_1^2)},~~~I_{2;\what{1}}=\int {d^D \ell\over (2\pi)^D}  {1\over (\ell^2-M_0^2)}.\label{oneloop-bub-3}\eea
If we collect these coefficients into a list ${\cal C}_2^{(r)}\equiv \left\{ C^{(r)}_{2\to 2},C^{(r)}_{2\to 2;\what{1}},C^{(r)}_{2\to 2;\what{0}}\right\}$, using the notations
\bea
f_1=K^2+M_0^2-M_1^2,~~~s_{00}=R^2,~~~s_{01}=R\cdot K,~~~s_{11}=K^2\co~~~~\label{oneloop-notation}\eea
we will have the following results:
\begin{itemize}
	
	\item {\bf For the rank $r=1$:}
	\bea {\cal C}_2^{(1)}=\left\{ {f_1s_{01}\over s_{11}},  -{s_{01}\over s_{11}}, {s_{01}\over s_{11}}\right\}\ed ~~~~~\label{C1-1}\eea

	\item {\bf For the rank $r=2$:}
	\bea {\cal C}_{2}^{(2)}& =& \left\{ {D f_1^2 s_{01}^2 \over (D-1)s_{11}^2}- { 4 M_0^2 s_{01}^2 +s_{00} (f_1^2-4 s_{11} M_0^2)\over (D-1)s_{11}}, ~~-{D f_1 s_{01}^2 \over (D-1)s_{11}^2}+ {s_{00} f_1\over (D-1) s_{11}},\right. \nn & & \left. {D f_1 s_{01}^2 \over (D-1)s_{11}^2}+ { 2(D-2) s_{01}^2 +s_{00} (s_{11}-M_0^2+M_1^2)\over (D-1)s_{11}}\right\}\ed~~~~~\label{C2-1}\eea
	For our application, we have separated terms according to the power of pole $s_{11}$.
	One crucial point is that we can write ${\cal C}_2^{(2)}$ as the combination of lower-rank coefficients and the contributions from lower topologies (tadpoles)
	\bea {\cal C}_2^{(2)}
	& = &  {D f_1 s_{01}\over (D-1) s_{11}}\mathcal{C}_2^{(1)}- { 4 M_0^2 s_{01}^2 +s_{00} (f_1^2-4 M_0^2s_{11})\over (D-1)s_{11}}{\cal C}_2^{(0)}\nn
	&&+\left\{ 0, {s_{00} f_1\over (D-1) s_{11}},\right. \left.  { 2(D-2) s_{01}^2 +s_{00} (s_{11}-M_0^2+M_1^2)\over (D-1)s_{11}}\right\}.~~~~~\label{C2-2}\eea
	\item {\bf For the rank $r=3$:} \\
	The three reduction coefficients  are
	\bea C^{(3)}_{2\to 2;\what{1}}& =& - {(D+2) f_1^2 s_{01}^3 \over (D-1) s_{11}^3} + {3 f_1^2 s_{00}s_{01}\over (D-1) s_{11}^2}+ {8 M_0^2 s_{01}^3 \over D s_{11}^2}-{12 M_0^2 s_{01} s_{00}\over D s_{11}}\co\nn
	C^{(3)}_{2\to 2;\what{0}}&= & { (s_{11}+M_1^2-M_0^2)^2 (-3 s_{00} s_{01} s_{11}+(D+2) s_{01}^3)\over (D-1) s_{11}^3}- {4 M_1^2 s_{01}( 2 s_{01}^2-3 s_{00} s_{11})\over D s_{11}^2}\co\nn
	& & + 6 s_{01} {(s_{11}+M_1^2-M_0^2)(s_{11} s_{00}- D s_{01}^2)\over (D-1) s_{11}^2}+12 s_{01}^2 {s_{01}\over s_{11}} \nn
	C^{(3)}_{2\to 2}&=& \frac{f_1 \left(s_{01}^3 \left((D+2) f_1^2-12 M_0^2 s_{11}\right)+3 s_{00} s_{11} s_{01} \left(4 M_0^2 s_{11}-f_1^2\right)\right)}{(D-1) s_{11}^3}\ed~~~~~~~~\label{C3-coeff}
	\eea
	By matching the terms with different powers of pole $s_{11}$, we can expand ${\cal C}_2^{(3)}$ as
	\begin{align}
		{\cal C}_2^{(3)}&= {(D+2) f_1 s_{01}\over D s_{11}} {\cal C}_2^{(2)}-\bigg [\frac{2 \left(f_1^2-4M_0^2s_{11}\right) s_{00}}{Ds_{11}}+\frac{8 M_0^2 s_{01}^2}{Ds_{11}}\bigg]{\cal C}_2^{(1)}\nn
		&\newline+\left\{0,-\frac{4 M_0^2 s_{00} s_{01}}{D s_{11}},\frac{4 \left(-M_0^2+2 M_1^2+s_{11}\right) s_{00} s_{01}}{D s_{11}}+\frac{4 (D-2) s_{01}^3}{D s_{11}}\right\}\ed~~~\label{A12}
	\end{align}
	One technical point of expansion \eref{A12} is that there are other different expansions, for example
	\bea  {\cal C}_2^{(3)}&=&  {(D+2) f_1 s_{01}\over D s_{11}} {\cal C}_2^{(2)}
	+ {-2 s_{00} (M_0^2-M_1^2)^2-8 M_0^2 s_{01}^2\over D s_{11}}{\cal C}_2^{(1)}\nn
	&&+ {2 s_{01}\over D s_{11}}\bigg\{ s_{00}(-f_1^2+f_1 (3 M_0^2+M_1^2)), s_{00}(s_{11}-4 M_0^2 -2 M_1^2), \nn
	&&\hspace{42pt}s_{00}(s_{11}+6 M_1^2)+2(D-2) s_{01}^2\bigg\}\ed~~~~~~~~\label{C3-2}\eea
	The difference between \eref{C3-2} and \eref{A12} is whether the coefficient of the bubble in the last term is zero or not.
	For simplicity of application, we prefer the form of \eref{A12}.
	The choice of bubble coefficient to be zero for the remaining term has fixed the expansion uniquely.

	\item {\bf For the rank $r=4$:}\\
	After some algebra, we have
	\begin{align}
	\mathcal{C}_2^{(4)}&=\frac{(D+4) f_1 s_{01}}{(D+1)s_{11}}{\cal C}_2^{(3)}-\bigg[\frac{3 (f_1^2-4M_0^2s_{11}) s_{00}}{(D+1)s_{11}}+\frac{12 M_0^2 s_{01}^2}{(D+1)s_{11}}\bigg]{\cal C}_2^{(2)}\nn
	&\newline+\bigg\{0,\frac{12 M_0^2 \left(M_0^2-M_1^2+s_{11}\right) s_{00}^2}{D (D+1) s_{11}},\frac{12 \left(-M_0^2+3 M_1^2+s_{11}\right) s_{00} s_{01}^2}{(D+1) s_{11}}+\frac{8 (D-2) s_{01}^4}{(D+1) s_{11}}\nn
	&\newline+ \frac{12 M_1^2 \left(-M_0^2+M_1^2+s_{11}\right) s_{00}^2}{D (D+1) s_{11}}\bigg\}\ed
	\end{align}
	\item {\bf For the rank $r=5$:}
	\begin{align}
	{\cal C}_2^{(5)}&=\frac{(D+6)f_1 s_{01}}{(D+2)s_{11}}{\cal C}_2^{(4)}-\bigg[\frac{4 (f_1^2-4M_0^2s_{11}) s_{00}}{(D+2) s_{11}}+\frac{16 M_0^2 s_{01}^2}{(D+2) s_{11}}\bigg]{\cal C}_2^{(3)}\nn
	&\newline+\bigg\{0,-\frac{48 M_0^4 s_{00}^2 s_{01}}{D(D+2)s_{11}},\frac{32 \left(-M_0^2+4 M_1^2+s_{11}\right) s_{00} s_{01}^3}{(D+2) s_{11}}+\frac{16 (D-2) s_{01}^5}{(D+2) s_{11}}\nn
	&\newline+\frac{48 M_1^2 \left(-2 M_0^2+3 M_1^2+2 s_{11}\right) s_{00}^2 s_{01}}{D (D+2) s_{11}}\bigg\}\ed
	\end{align}

\end{itemize}

From these examples, one can observe the general
expansion  for any rank $r>0$ to be
\begin{align}
I_2^{(r)}&=\frac{{(D+2(r-2))f_1s_{01}}I_2^{(r-1)}-(r-1)\big(4M_0^2s_{01}^2+(f_1^2-4M_0^2s_{11})s_{00}\big )I_2^{(r-2)}+{\cal I}_{Tad}^{(r)}}{(D+r-3)s_{11}}~~~\label{A16}
\end{align}
where ${\cal I}_{Tad}^{(r)}$  is the contribution of tadpoles containing no poles of $s_{11}$. Relation \eref{A16} is one key relation used in the paper, so it is desirable to give proof.
Note that reduction coefficients are completely determined by recursion relations produced by $\cal D$-type operators and a $\cal T$-type operator. We can prove \eref{A16} by arguing that it is a solution to all $\cal D$-type recursions and  $\cal T$-type recursion relation in \eref{Relations}. We will do the proof recursively: assuming \eref{A16} is valid for all $r<r_0$, then it is also true for $r_0$.

Let us start from  $\cal D$-type recursion relations, i.e., the differential equation generated by ${\cal D}_1$ for the bubble.
Acting with $K\d_R$ on both sides of \eref{A12} and using $2K\cdot \ell=D_0-D_1+f_1$. The LHS is
\bea
{\cal D}_1 I_{2}^{(r_0)}=r_0f_1I_{2}^{(r_0-1)}+r_0I_{2;\what{0}-\what{1}}^{(r_0-1)}=r_0f_1I_{2}^{(r_0-1)}+O(\text{tad})\ed\label{A-17}
\eea
The term $r_0I_{2;\what{0}-\what{1}}^{(r-1)}$ can be reduced to scalar tadpole without any poles from  Gram determent $s_{11}$, and for simplicity, we just denote the contribution of tadpole topology as $O(\text{tad}) $. The RHS becomes
\begin{align}
&\frac{1}{(D+r_0-3)s_{11}}\Big[(D+2(r_0-2))f_1(s_{11}I_2^{(r_0-1)}+(r_0-1)s_{01}f_1I_2^{(r_0-2)})\nn
&-(r_0-1)\big(2f_1^2s_{01}\big )I_2^{(r_0-2)}-(r_0-1)\big(4M_0^2s_{01}^2+(f_1^2-4M_0^2s_{11})s_{00}\big )(r_0-2)f_1I_2^{(r_0-3)}+O(\text{tad})\Big]\co\label{A-18}
\end{align}
To require  \eref{A-18} be equal to \eref{A-17}, one need following equation
\bea
I_2^{(r_0-1)}=\frac{(D+2(r_0-3))f_1s_{01}I_2^{(r_0-2)}-(r_0-2)\big(4M_0^2s_{01}^2+(f_1^2-4M_0^2s_{11})s_{00}\big )I_2^{(r_0-3)}+{O(\text{tad})}}{(D+r_0-4)s_{11}}\co\nn
\eea
which is just \eref{A16} with $r=r_0-1$.

Then we consider the $\cal T$-type recursion. Acting with ${\cal T}$ on the LHS of \eref{A16}, we get
\bea
{\cal T}I_2^{(r_0)}=4r_0(r_0-1)M_0^2I_2^{(r_0-2)}+O(\text{tad})\co \label{A-20}
\eea
while acting on the RHS we get
\bea
&&\frac{1}{(D+r_0-3)s_{11}}\Bigg[(D+2(r_0-2))f_1\left[2(r_0-1)f_1I_2^{(r_0-2)}+4s_{01}(r_0-1)(r_0-2)M_0^2I_2^{(r_0-3)}\right]\nn
&&-(r_0-1)\big(4M_0^2\left[2s_{11}I_2^{(r_0-2)}+4(r_0-2)s_{01}f_1I_2^{(r_0-3)}+4(r_0-2)(r_0-3)s_{01}^2M_0^2I_2^{(r_0-4)}\right]\nn
&&+(f_1^2-4M_0^2s_{11})\left[2(D+2r_0-4)I_2^{(r_0-2)}+4(r_0-2) (r_0-3)M_0^2s_{00}I^{(r_0-4)})\right]\big )+O(\text{tad})\Bigg]\ed\nn \label{A-21}
\eea
One can prove  \eref{A-20} is equal to  \eref{A-21} by employing the recursion relation \eref{A16} for rank $r_0-2$.

One crucial point of the above proof is that, in principle, one can establish recursion relations of ${\cal I}_{Tad}^{(r)}$ for general $r$. Thus we can solve it and get the complete relation \eref{A16} with explicit expression for ${\cal I}_{Tad}^{(r)}$.
This result will give another recursive way to analytically compute reduction coefficients, which may be more efficient
than the one given in \cite{Hu:2021nia}. Furthermore, knowing ${\cal I}_{Tad}^{(r)}$, one may write down
the analytic expression of $\vec{\beta}_{\ell_1;i_1,i_2,j}^{(r_1,r_2)}$ in \eref{recur-l1} and its dual form.
Then we can get the algebraic recursion relation for
all coefficients for any rank level, just like what has been done
in \cite{Feng:2021enk,Hu:2021nia}. It is obvious that with higher and
higher rank levels, all available programs, like FIRE, Kira, Reduce \cite{vonManteuffel:2012np,Lee:2013mka,Smirnov:2019qkx,Maierhofer:2017gsa} will become harder and harder  to handle analytically. However, the method proposed in this paper can still work with fewer efforts.

\section{Reduction of lower topologies $I^{(r_1,r_2)}_{1,1,1;\what{i}}$}
\label{lower-toplogy}
Here we give the reduction results for integrals got by removing the $i$-th propagator in the tensor sunsets, i.e., $I^{(r_1,r_2)}_{1,1,1;\what{i}}$.
We denote
\begin{align}
I^{(r_1,r_2)}_{1,1,1;\what{i}}={\bf C}^{(r_1,r_2)}_{ \what{i}}{\bf J}\ed
~~~\label{B1}
\end{align}
The reduction coefficients ${\bf C}^{(r_1,r_2)}_{ \what{i}}$ can be  obtained using  the results \eref{oneloop-tad-3}. Among the three lower
topologies, $I^{(r_1,r_2)}_{1,1,1;\what{3}}$ is trivial, which is given by
\begin{align}
I^{(r_1,r_2)}_{1,1,1;\what{3}}&=\int {d\ell_1d\ell_2(2\ell_1\cdot R_1)^{r_1}(2\ell_2\cdot R_2)^{r_2}\over (\ell_2^2-M_2^2)(\ell_2^2-M_2^2)}\nn %
&=\gamma(r_1,M_1)\gamma(r_2,M_2)(R_1^2)^{r_1\over 2}(R_2^2)^{r_2\over 2}\int {d\ell_1 d\ell_2\over (\ell_1^2-M_1^2)(\ell_2^2-M_2^2)}\nn
&=\gamma(r_1,M_1)\gamma(r_2,M_2)(R_1^2)^{r_1\over 2}(R_2^2)^{r_2\over 2}~J_7\co ~~~\label{B2}
\end{align}
from which we can easily read out ${\bf C}^{(r_1,r_2)}_{ \what{i}}$
in \eref{B1}.
For $I^{(r_1,r_2)}_{1,1,1;\what{1}}$, because the subtlety discussed in
\eref{3-45}, we need to shift $\ell_1$ in the numerator. The computation details are:
\allowdisplaybreaks
\begin{align}
I^{(r_1,r_2)}_{1,1,1;\what{1}}&=\int {d\ell_1d\ell_2(2\ell_1\cdot R_1)^{r_1}(2\ell_2\cdot R_2)^{r_2}\over (\ell_2^2-M_2^2)[(\ell_1+\ell_2-K)^2-M_3^2]}\nn
&\xlongequal{\ell_1 \to \ell_1-\ell_2+K}\int {d\ell_1d\ell_2(2\ell_1\cdot R_1-2\ell_2\cdot R_1+2 K\cdot R_1)^{r_1}(2\ell_2\cdot R_2)^{r_2}\over (\ell_2^2-M_2^2)(\ell_1^2-M_3^2)}\nn
&=\sum_{i=0}^{r_1} \binom{r_1}{i}(2R_1\cdot K)^{r_1-i}\int {d\ell_1d\ell_2(2\ell_1\cdot R_1-2\ell_2\cdot R_1)^{i}(2\ell_2\cdot R_2)^{r_2}\over (\ell_2^2-M_2^2)(\ell_1^2-M_3^2)}\nn
&=\sum_{i=0}^{r_1}\sum_{j=0}^i\binom{r_1}{i}\binom{i}{j} (2R_1\cdot K)^{r_1-i}\int {d\ell_1d\ell_2(2\ell_1\cdot R_1)^{i-j}(-2\ell_2\cdot R_1)^{j}(2\ell_2\cdot R_2)^{r_2}\over (\ell_2^2-M_2^2)(\ell_1^2-M_3^2)}\nn
&=\sum_{i=0}^{r_1}\sum_{j=0}^i\binom{r_1}{i}\binom{i}{j} (2R_1\cdot K)^{r_1-i}\gamma(i-j,M_3)(R_1^2)^{i-j\over 2}\nn
&\newline
\times (-1)^j\gamma(r_2+j,M_2){r_2!(R_1\cdot \d_{R_2})^j\over (r_2+j)!}(R_2^2)^{r_2+j\over 2}\int {1\over (\ell_2^2-M_2^2)(\ell_1^2-M_3^2)}\nn
&=\sum_{i=0}^{r_1}\sum_{j=0}^i\binom{r_1}{i}\binom{i}{j}\gamma(i-j,M_3) \gamma(r_2+j,M_2){(-1)^jr_2!\over (r_2+j)!}(2R_1\cdot K)^{r_1-i}(R_1^2)^{i-j\over 2}\nn
&\newline
\times(R_1\cdot \d_{R_2})^j(R_2^2)^{r_2+j\over 2}\int {d\ell_1 d\ell_2\over (\ell_2^2-M_2^2)(\ell_1^2-M_3^2)}\ed~~~\label{B3}
\end{align}
Similarly, we have
\begin{align}
I^{(r_1,r_2)}_{1,1,1;\what{2}}&=\sum_{i=0}^{r_2}\sum_{j=0}^i\binom{r_2}{i}\binom{i}{j}\gamma(i-j,M_3) \gamma(r_1+j,M_1){(-1)^jr_1!\over (r_1+j)!}(2R_2\cdot K)^{r_2-i}(R_2^2)^{i-j\over 2}\nn
&\newline
\times(R_2\cdot \d_{R_1})^j(R_1^2)^{r_1+j\over 2}\int {d\ell_1 d\ell_2\over (\ell_1^2-M_1^2)(\ell_2^2-M_3^2)}\ed~~~\label{B4}
\end{align}

\section{Lower rank/topology terms in $\cal D$-type relation}
\label{apdix:lowerterm}
To write down ${\cal D}$-type recursion relation for given $r_1,r_2$,
we need calculate  ${\cal B}^{(r_1,r_2)}_{\ell_{1}},{\cal B}^{(r_1,r_2)}_{\ell_{2}}$ defined in \eref{3.31}.
Here we list some results. For the rank  $(1,r_2)$ and $(r_1,1)$, the general forms  are
	\begin{align}
	{\cal B}^{(1,r_2)}_{\ell_{1}}&=(D-2)\Big[(s_{11}+M_2^2){\bf C}^{(1,r_2)}+{\bf C}^{(1,r_2)}_{\what{2}}+{s_{01}{\cal D}_{10'}\over r_2+1}{\bf C}^{(0,r_2+1)}\nn
	&\newline-s_{01}\left({\bf C}^{(0,r_2)}_{\what{1}+\what{2}-\what{3}}+f_{12}{\bf C}^{(0,r_2)}\right)+{{\cal T}_{00'}\over 2(r_2+1)}\left({\bf C}^{(0,r_2+1)}_{\what{1}+\what{2}-\what{3}}+f_{12}{\bf C}^{(0,r_2+1)}\right)\Big]\nn
	{\cal B}^{(r_1,1)}_{\ell_{2}}&=(D-2)\Big[(s_{11}+M_1^2){\bf C}^{(r_1,1)}+{\bf C}^{(r_1,1)}_{\what{1}}+{s_{0'1}{\cal D}_{10}\over r_1+1}{\bf C}^{(r_1+1,0)}\nn
	&\newline-s_{0'1}\left({\bf C}^{(r_1,0)}_{\what{1}+\what{2}-\what{3}}+f_{12}{\bf C}^{(0,r_2)}\right)+{{\cal T}_{0'0}\over 2(r_1+1)}\left({\bf C}^{(r_1+1,0)}_{\what{1}+\what{2}-\what{3}}+f_{12}{\bf C}^{(r_1+1,0)}\right)\Big]\ed
	\end{align}
One can also write down expression for $(2,r_2)$, $(3,r_2)$ etc. The expression will become longer and longer. Since in examples the rank
level is up to four,  we just list the terms we used in the examples
	\allowdisplaybreaks
\begin{align}
\allowdisplaybreaks
{\cal B}^{(1,0)}_{\ell_{1}}&=-\frac{1}{2} (D-2) s_{01} \Big[2 f_{12} \vec{\a} _{0,0,0}^{(0,0)}-f_{12} \vec{\a} _{0,1,0}^{(0,1)}+2 \vec{\a} _{0,0,0;\what{2}}^{(0,0)}-\vec{\a} _{0,1,0;\what{2}}^{(0,1)}-2 \vec{\a} _{1,0,0;\what{2}}^{(1,0)}\nn
&\newline-2 M_2^2 \vec{\a} _{1,0,0}^{(1,0)}-2 s_{11} \vec{\a} _{0,1,0}^{(0,1)}-2 s_{11} \vec{\a} _{1,0,0}^{(1,0)}+2 \vec{\delta}_5-2 \vec{\delta}_7\Big]\co\nn
{\cal B}^{(1,1)}_{\ell_{1}}&=\frac{1}{2} (D-2) s_{00'} \left[f_{12} \vec{\a} _{0,0,0}^{(0,2)}+\vec{\a} _{0,0,0;\what{2}}^{(0,2)}+2 \vec{\a} _{0,0,1;\what{2}}^{(1,1)}+2 \left(M_2^2+s_{11}\right) \vec{\a} _{0,0,1}^{(1,1)}+{4\over D} \left(\vec{\delta}_5-\vec{\delta}_7\right) M_2^2\right]\nn
&+\frac{1}{2} (D-2) s_{01} s_{0'1} \Big[-2 f_{12} \vec{\a} _{0,1,0}^{(0,1)}+f_{12} \vec{\a} _{0,2,0}^{(0,2)}-2 \vec{\a} _{0,1,0;\what{2}}^{(0,1)}+\vec{\a} _{0,2,0;\what{2}}^{(0,2)}+2 \vec{\a} _{1,1,0;\what{2}}^{(1,1)}\nn
&\newline \hspace{100pt}+2 M_2^2 \vec{\a} _{1,1,0}^{(1,1)}+2 s_{11} \vec{\a} _{0,2,0}^{(0,2)}+2 s_{11} \vec{\a} _{1,1,0}^{(1,1)}+2 \vec{\a} _{0,0,0}^{(0,2)}\Big]\co\nn
{\cal B}^{(1,2)}_{\ell_{1}}&= \frac{1}{6 D}(D-2) s_{0'0'}s_{01} \bigg[-6 D f_{12} \vec{\a} _{0,0,0}^{(0,2)}+D f_{12} \vec{\a} _{0,1,0}^{(0,3)}-6 D \vec{\a} _{0,0,0;\what{2}}^{(0,2)}+D \vec{\a} _{0,1,0;\what{2}}^{(0,3)}\nn
&\newline+6 D \vec{\a} _{1,0,0;\what{2}}^{(1,2)}+6 D M_2^2 \vec{\a} _{1,0,0}^{(1,2)}+2 D s_{11} \vec{\a} _{0,1,0}^{(0,3)}+6 D s_{11} \vec{\a} _{1,0,0}^{(1,2)}-24 \vec{\delta}_5 M_2^2+24 \vec{\delta}_7 M_2^2\bigg]\nn
&+\frac{1}{6} (D-2) s_{0'1}^2s_{01} \bigg[-6 f_{12} \vec{\a} _{0,2,0}^{(0,2)}+3 f_{12} \vec{\a} _{0,3,0}^{(0,3)}-6 \vec{\a} _{0,2,0;\what{2}}^{(0,2)}+3 \vec{\a} _{0,3,0;\what{2}}^{(0,3)}\nn
&\newline+6 \vec{\a} _{1,2,0;\what{2}}^{(1,2)}+6 M_2^2 \vec{\a} _{1,2,0}^{(1,2)}+6 s_{11} \vec{\a} _{0,3,0}^{(0,3)}+6 s_{11} \vec{\a} _{1,2,0}^{(1,2)}+4 \vec{\a} _{0,1,0}^{(0,3)}\bigg]\nn
&+\frac{1}{3} (D-2) s_{00'} s_{0'1} \left(f_{12} \vec{\a} _{0,1,0}^{(0,3)}+3 \left(\vec{\a} _{0,1,1;\what{2}}^{(1,2)}+\left(M_2^2+s_{11}\right) \vec{\a} _{0,1,1}^{(1,2)}\right)+\vec{\a} _{0,1,0;\what{2}}^{(0,3)}\right)\co\nn
{\cal B}^{(2,0)}_{\ell_{1}}&=s_{01}^2 \bigg[-D f_{12} \vec{\a} _{1,0,0}^{(1,0)}+\frac{1}{2} D f_{12} \vec{\a} _{1,1,0}^{(1,1)}-D \vec{\a} _{1,0,0;\what{2}}^{(1,0)}+\frac{1}{2} D \vec{\a} _{1,1,0;\what{2}}^{(1,1)}+(D-1) \vec{\a} _{2,0,0;\what{2}}^{(2,0)}+D s_{11} \vec{\a} _{1,1,0}^{(1,1)}\nn
&\newline+(D-1) \left(M_2^2+s_{11}\right) \vec{\a} _{2,0,0}^{(2,0)}+D \vec{\a} _{0,0,1}^{(1,1)}+4 M_1^2 \vec{\a} _{0,0,0}^{(0,0)}-4 M_1^2 \vec{\a} _{0,1,0}^{(0,1)}+M_1^2 \vec{\a} _{0,2,0}^{(0,2)}+4 \vec{\delta}_5-2 \vec{\delta}_5 D\bigg]\nn
&+s_{00} \bigg[\frac{1}{2} D \left(f_{12} \vec{\a} _{0,0,1}^{(1,1)}+\vec{\a} _{0,0,1;\what{2}}^{(1,1)}\right)+\frac{1}{4} D \vec{\a} _{0,0,0;\what{2}}^{(0,2)}+(D-1) \vec{\a} _{0,0,0;\what{2}}^{(2,0)}+(D-1) \left(M_2^2+s_{11}\right) \vec{\a} _{0,0,0}^{(2,0)}\nn
&\newline +2 f_{12} \vec{\a} _{0,0,0;\what{2}}^{(0,0)}-2 s_{11} \left(f_{12}-2 M_1^2\right) \vec{\a} _{0,1,0}^{(0,1)}+f_{12}^2 \vec{\a} _{0,0,0}^{(0,0)}-4 M_1^2 \vec{\a} _{0,0,0;\what{2}}^{(0,0)}-M_2^2 \vec{\a} _{0,0,0;\what{2}}^{(0,0)}-2 s_{11} \vec{\a} _{0,1,0;\what{2}}^{(0,1)}\nn
&\newline+\frac{1}{4} s_{11} \vec{\a} _{0,2,0;\what{2}}^{(0,2)}-4 M_1^2 s_{11} \vec{\a} _{0,0,0}^{(0,0)}-4 M_1^2 M_2^2 \vec{\a} _{0,0,0}^{(0,0)}+M_1^2 \vec{\a} _{0,0,0}^{(0,2)}+s_{11} \left(s_{11} \vec{\a} _{0,2,0}^{(0,2)}+\vec{\a} _{0,0,0}^{(0,2)}\right)\nn
&\newline+\frac{4 \vec{\delta}_5 M_2^2}{D}+\vec{\delta}_5 M_1^2-3 \vec{\delta}_5 M_2^2-\vec{\delta}_5 M_3^2-\vec{\delta}_7 M_1^2-\vec{\delta}_7 M_2^2+\vec{\delta}_7 M_3^2-\vec{\delta}_5 s_{11}-\vec{\delta}_7 s_{11}\bigg]\co\nn
{\cal B}^{(2,1)}_{\ell_{1}}
	&=s_{0'1} s_{01}^2 \Big[4 M_1^2 \vec{\a} _{0,1,0}^{(0,1)}-4 M_1^2 \vec{\a} _{0,2,0}^{(0,2)}+M_1^2 \vec{\a} _{0,3,0}^{(0,3)}-D f_{12} \vec{\a} _{1,1,0}^{(1,1)}-D \vec{\a} _{1,1,0;\what{2}}^{(1,1)}+\frac{1}{2} D \vec{\a} _{0,1,1}^{(1,2)}+D \vec{\a} _{1,0,0}^{(1,2)}\nn
	&\newline +\frac{1}{2} D f_{12} \vec{\a} _{1,2,0}^{(1,2)}+D s_{11} \vec{\a} _{1,2,0}^{(1,2)}+\frac{1}{2} D \vec{\a} _{1,2,0;\what{2}}^{(1,2)}+(D-1) \left(M_2^2+s_{11}\right) \vec{\a} _{2,1,0}^{(2,1)}+(D-1) \vec{\a} _{2,1,0;\what{2}}^{(2,1)}\Big]\nn
	&+s_{00'}s_{01} \Big[-4 M_1^2 \vec{\a} _{0,0,0}^{(0,2)}+\frac{2}{3} M_1^2 \vec{\a} _{0,1,0}^{(0,3)}-D f_{12} \vec{\a} _{0,0,1}^{(1,1)}-D \vec{\a} _{0,0,1;\what{2}}^{(1,1)}+\frac{1}{4} D f_{12} \vec{\a} _{0,1,1}^{(1,2)}+\frac{1}{2} D s_{11} \vec{\a} _{0,1,1}^{(1,2)}\nn
	&\newline+\frac{1}{2} D f_{12} \vec{\a} _{1,0,0}^{(1,2)}+\frac{1}{4} D \vec{\a} _{0,1,1;\what{2}}^{(1,2)}+\frac{1}{2} D \vec{\a} _{1,0,0;\what{2}}^{(1,2)}+(D-1) \left(M_2^2+s_{11}\right) \vec{\a} _{1,0,1}^{(2,1)}+(D-1) \vec{\a} _{1,0,1;\what{2}}^{(2,1)}\nn
	&\newline-\frac{16 M_2^2 \vec{\delta}_5}{D}+8 M_2^2 \vec{\delta}_5\Big]\nn
	&+s_{00} s_{0'1} \Big[f_{12}^2 \vec{\a} _{0,1,0}^{(0,1)}-4 M_1^2 M_2^2 \vec{\a} _{0,1,0}^{(0,1)}-4 M_1^2 s_{11} \vec{\a} _{0,1,0}^{(0,1)}+\frac{1}{3} M_1^2 \vec{\a} _{0,1,0}^{(0,3)}+s_{11} \vec{\a} _{0,1,0}^{(0,3)}-4 M_1^2 \vec{\a} _{0,1,0;\what{2}}^{(0,1)}\nn
	&\newline+(D-1) \left(M_2^2+s_{11}\right) \vec{\a} _{0,1,0}^{(2,1)}-M_2^2 \vec{\a} _{0,1,0;\what{2}}^{(0,1)}+2 f_{12} \vec{\a} _{0,1,0;\what{2}}^{(0,1)}-2 \left(f_{12}-2 M_1^2\right) \vec{\a} _{0,0,0}^{(0,2)}\nn
	&\newline-2 \left(f_{12}-2 M_1^2\right) s_{11} \vec{\a} _{0,2,0}^{(0,2)}-2 \vec{\a} _{0,0,0;\what{2}}^{(0,2)}-2 s_{11} \vec{\a} _{0,2,0;\what{2}}^{(0,2)}+s_{11}^2 \vec{\a} _{0,3,0}^{(0,3)}+\frac{1}{12} D \vec{\a} _{0,1,0;\what{2}}^{(0,3)}+\frac{1}{6} \vec{\a} _{0,1,0;\what{2}}^{(0,3)}\nn
	&\newline+\frac{1}{4} s_{11} \vec{\a} _{0,3,0;\what{2}}^{(0,3)}+\frac{1}{4} D f_{12} \vec{\a} _{0,1,1}^{(1,2)}+\frac{1}{4} D \vec{\a} _{0,1,1;\what{2}}^{(1,2)}+(D-1) \vec{\a} _{0,1,0;\what{2}}^{(2,1)}+\frac{4 M_2^2 \vec{\delta}_5}{D}+\frac{4 M_2^2 \vec{\delta}_7}{D}\Big]\co\nn
   {\cal B}^{(3,0)}_{\ell_{1}}&=s_{01}^3 \bigg[-(D+2) f_{12} \vec{\a} _{2,0,0}^{(2,0)}+\frac{1}{2} (D+2) f_{12} \vec{\a} _{2,1,0}^{(2,1)}-(D+2) \vec{\a} _{2,0,0;\what{2}}^{(2,0)}+\frac{1}{2} (D+2) \vec{\a} _{2,1,0;\what{2}}^{(2,1)}\nn
&\newline +D \vec{\a} _{3,0,0;\what{2}}^{(3,0)}+D \left(M_2^2+s_{11}\right) \vec{\a} _{3,0,0}^{(3,0)}+(D+2) s_{11} \vec{\a} _{2,1,0}^{(2,1)}+(D+2) \vec{\a} _{1,0,1}^{(2,1)}+8 M_1^2 \vec{\a} _{1,0,0}^{(1,0)}\nn
&\newline-8 M_1^2 \vec{\a} _{1,1,0}^{(1,1)}+2 M_1^2 \vec{\a} _{1,2,0}^{(1,2)}+8 \vec{\delta}_5-4 \vec{\delta}_5 D\bigg]\nn
&+\frac{1}{2} s_{00} s_{01} \bigg[-2 (D+2) f_{12} \vec{\a} _{0,0,0}^{(2,0)}+(D+2) f_{12} \vec{\a} _{0,1,0}^{(2,1)}+(D+2) f_{12} \vec{\a} _{1,0,1}^{(2,1)}+D \vec{\a} _{1,0,0;\what{2}}^{(1,2)}\nn
&\newline-2 (D+2) \vec{\a} _{0,0,0;\what{2}}^{(2,0)}+(D+2) \vec{\a} _{0,1,0;\what{2}}^{(2,1)}+(D+2) \vec{\a} _{1,0,1;\what{2}}^{(2,1)}+2 D \vec{\a} _{1,0,0;\what{2}}^{(3,0)}+2 D \left(M_2^2+s_{11}\right) \vec{\a} _{1,0,0}^{(3,0)}\nn
&\newline+2 (D+2) s_{11} \vec{\a} _{0,1,0}^{(2,1)}+8 f_{12} \vec{\a} _{1,0,0;\what{2}}^{(1,0)}-8 s_{11} \left(f_{12}-2 M_1^2\right) \vec{\a} _{1,1,0}^{(1,1)}-8 \left(f_{12}-2 M_1^2\right) \vec{\a} _{0,0,1}^{(1,1)}\nn
&\newline+4 f_{12}^2 \vec{\a} _{1,0,0}^{(1,0)}-16 M_1^2 \vec{\a} _{1,0,0;\what{2}}^{(1,0)}-4 M_2^2 \vec{\a} _{1,0,0;\what{2}}^{(1,0)}-8 s_{11} \vec{\a} _{1,1,0;\what{2}}^{(1,1)}+s_{11} \vec{\a} _{1,2,0;\what{2}}^{(1,2)}-8 \vec{\a} _{0,0,1;\what{2}}^{(1,1)}+\vec{\a} _{0,1,1;\what{2}}^{(1,2)}\nn
&\newline-16 M_1^2 s_{11} \vec{\a} _{1,0,0}^{(1,0)}-16 M_1^2 M_2^2 \vec{\a} _{1,0,0}^{(1,0)}+4 M_1^2 \vec{\a} _{1,0,0}^{(1,2)}-16 M_1^2 \vec{\a} _{0,0,1}^{(1,1)}+4 M_1^2 \vec{\a} _{0,1,1}^{(1,2)}+4 s_{11} \vec{\a} _{1,0,0}^{(1,2)}\nn
&\newline+4 s_{11} \vec{\a} _{0,1,1}^{(1,2)}+4 s_{11}^2 \vec{\a} _{1,2,0}^{(1,2)}+\frac{32 \vec{\delta}_5 M_2^2}{D}+8 \vec{\delta}_5 M_1^2-32 \vec{\delta}_5 M_2^2-16 \vec{\delta}_5 M_3^2+8 \vec{\delta}_7 M_1^2-8 \vec{\delta}_5 s_{11}\bigg]\ed
\end{align}

For the dual cases ${\cal B}_{\ell_2}^{(a,b)}$, one can obtain them from above results with proper replacements.

\section{Transfer matrix between tensor MIs and scalar MIs}
\label{apdix:transmatrix}
In section \ref{subsec:MIs}, we have discussed the choice of master basis. The basis chose by {\sc FIRE6} is
\bea
{\bf F}=\left\{I_{2,1,1},I_{1,2,1},I_{1,1,2},I_{1,1,1},I_{1,1,0},I_{1,0,1},I_{0,1,1}\right\}\co
\eea
and the basis used in this paper is
\bea
&&J_1=\int d \ell_{1,2} {1\over D_1D_2D_3},
J_2=\int d \ell_{1,2} {2\ell_1\cdot K\over D_1D_2D_3},
J_3=\int d \ell_{1,2} {2\ell_2\cdot K\over D_1D_2D_3},\nn
&&J_4=\int d \ell_{1,2} {(2\ell_1\cdot K)(2\ell_2 \cdot K)\over D_1D_2D_3},J_5=\int  {d\ell_{1,2}\over D_2D_3},J_6=\int  {d\ell_{1,2}\over D_1D_3},J_7=\int  {d\ell_{1,2}\over D_1D_2}\ed
\eea
The transform matrix ${\bf J}=T{\bf F}$ is
	\begin{small}
	\allowdisplaybreaks
    \begin{align}
	T_{1,1\sim7}&=\{0,0,0,1,0,0,0\}\co\nn
	T_{2,1\sim 7}&=\bigg\{\frac{4 M_1 \left(M_1-s_{11}\right)}{3 (D-2)},\frac{2 M_2 \left(3 M_1-M_2-3
		M_3+s_{11}\right)}{3 (D-2)},\frac{2 M_3 \left(3 M_1-3 M_2-M_3+s_{11}\right)}{3
		(D-2)},\nn
	&\frac{2 \left(-2 (D-3) M_1+(D-3) M_2+D M_3+D s_{11}-3 M_3-2 s_{11}\right)}{3
		(D-2)},-\frac{1}{3},-\frac{1}{3},\frac{2}{3}\bigg\}\co\nn
	T_{3,1\sim 7}&=\bigg\{\frac{2 M_1 \left(-M_1+3 M_2-3 M_3+s_{11}\right)}{3 (D-2)},\frac{4 M_2
		\left(M_2-s_{11}\right)}{3 (D-2)},\frac{2 M_3 \left(-3 M_1+3 M_2-M_3+s_{11}\right)}{3
		(D-2)},\nn&\frac{2 \left((D-3) M_1-2 (D-3) M_2+D M_3+D s_{11}-3 M_3-2 s_{11}\right)}{3
		(D-2)},-\frac{1}{3},\frac{2}{3},-\frac{1}{3}\bigg\}\co\nn
	T_{4,1}&=-\frac{4 (2 D-3) M_1^2 s_{11}^2}{3 (D-2) (3 D-4)}+\frac{4 M_1^2 s_{11} \left((3 D-4) M_1^2+(4 D-7) M_2^2+(1-2 D) M_3^2\right)}{3 (D-2) (3 D-4)}\nn&\newline-\frac{4 M_1^4 \left((D-1) M_1^2-5 (D-1) M_2^2+(7 D-11) M_3^2\right)}{3 (D-2) (3 D-4)}\co\nn
	T_{4,2}&=\frac{4 (3-2 D) M_2^2 s_{11}^2}{3 (D-2) (3 D-4)}+\frac{4 M_2^2 s_{11} \left((4 D-7) M_1^2+(3 D-4) M_2^2+(1-2 D) M_3^2\right)}{3 (D-2) (3 D-4)}\nn&\newline-\frac{4 M_2^4 \left(-5 (D-1) M_1^2+(D-1) M_2^2+(7 D-11) M_3^2\right)}{3 (D-2) (3 D-4)}\co\nn
	T_{4,3}&=\frac{2 (5 D-6) M_3^2 s_{11}^2}{3 (D-2) (3 D-4)}-\frac{4 M_3^2 s_{11} \left((7-4 D) (M_1^2+M_2^2)+(3 D-4) M_3^2\right)}{3 (D-2) (3 D-4)}+\frac{2 M_3^2(12-9 D) (M_1^4+M_2^4)}{3 (D-2) (3 D-4)}\nn&\newline+\frac{2 M_3^2 \left(2 M_1^2 \left(3 (3 D-4) M_2^2+(7-4 D) M_3^2\right)+(D-2) M_3^4+2 (7-4 D) M_2^2 M_3^2\right)}{3 (D-2) (3 D-4)}\co\nn
	T_{4,4}&=\frac{(6-4 D) s_{11}^2}{12-9 D}-\frac{2 s_{11} \left(\left(-7 D^2+26 D-20\right) M_3^2+\left(8 D^2-33 D+32\right) \left(M_1^2+M_2^2\right)\right)}{3 (D-2) (3 D-4)}\nn&\newline+\frac{2 (D-3) \left(-(D-2) M_3^4+(7 D-12) (M_1^2+M_2^2) M_3^2+2 (D-1) \left(M_1^4+M_2^4\right)- 8 (D-1) M_1^2M_2^2\right)}{3 (D-2) (3 D-4)}\co\nn
	T_{4,5}&=\frac{(6-4 D) s_{11}}{12-9 D}-\frac{2 \left((D-1) M_1^2+(D-1) M_2^2+(D-2) M_3^2\right)}{9 D-12}\co\nn
	T_{4,6}&=\frac{(7 D-10) M_1^2+(8-5 D) M_2^2+(D-2) M_3^2}{9 D-12}+\frac{(6-5 D) s_{11}}{9 D-12}\co\nn
	T_{4,7}&=\frac{(8-5 D) M_1^2+(7 D-10) M_2^2+(D-2) M_3^2}{9 D-12}+\frac{(6-5 D) s_{11}}{9 D-12}\co\nn
	T_{5,1\sim 7}&=\{0,0,0,0,0,0,1\}\co\nn
	T_{6,1\sim 7}&=\{0,0,0,0,0,1,0\}\co\nn
	T_{7,1\sim 7}&=\{0,0,0,0,1,0,0\}\ed
	\end{align}
	\end{small}
One can check that with general masses and $K$, it is non-degenerate.

\section{Results}
\label{apdix:results}
Here we collect the reduction results from rank level one to three for reference. It is too long to write in text form for rank level four, so we collect them in an attached Mathematica file.
\subsection*{$\bullet$ rank $r_1+r_2=1$}
\bea
\vec{\a}^{(1,0)}_{1,0,0}={1\over s_{11}}\left\{0,1,0,0,0,0,0\right\}\co~~~ \vec{\a}^{(0,1)}_{0,1,0}={1\over s_{11}}\left\{0,0,1,0,0,0,0\right\}\ed
\eea
\subsection*{$\bullet$ rank $r_1+r_2=2$}
	\begin{align}
	{\vec{\a}}^{(1,1)}_{0,0,1}&={1\over D-1}\left\{-2 f_{12},2,2,-\frac{1}{s_{11}},-2,-2,2\right\}\co\nn
	{\vec{\a}}^{(1,1)}_{1,1,0}&={1\over (D-1)s_{11}}\left\{2 f_{12},-2,-2,\frac{D}{s_{11}},2,2,-2\right\}\co\nn
	\vec{\a}_{0,0,0}^{(2,0)}&={1\over D-1}\left\{2 \left(f_{12}+2 M_1^2\right),-\frac{f_{12}}{s_{11}}-2,-\frac{2 \left(M_1^2+s_{11}\right)}{s_{11}},\frac{2}{s_{11}},4,2,-2\right\}\co\nn
	\vec{\a}_{2,0,0}^{(2,0)}&={-1\over (D-1)s_{11}}\left\{2 \left(D f_{12}+2 M_1^2\right),\frac{-D \left(f_{12}+2 s_{11}\right)}{s_{11}},\frac{-2 D \left(M_1^2+s_{11}\right)}{s_{11}},\frac{2 D}{s_{11}},4,2 D,-2 D\right\}\co\nn
	\vec{\a}_{0,0,0}^{(0,2)}&={1\over D-1}\left\{2 \left(f_{12}+2 M_2^2\right),-\frac{2 \left(M_2^2+s_{11}\right)}{s_{11}},-\frac{f_{12}}{s_{11}}-2,\frac{2}{s_{11}},2,4,-2\right\}\co\nn
	\vec{\a}_{0,2,0}^{(0,2)}&={-1\over (D-1)s_{11}}\left\{2 \left(D f_{12}+2 M_2^2\right),\frac{-2 D \left(M_2^2+s_{11}\right)}{s_{11}},\frac{-D \left(f_{12}+2 s_{11}\right)}{s_{11}},\frac{2 D}{s_{11}},2 D,4,-2 D\right\}\ed\nn
	\end{align}
\subsection*{$\bullet$ rank $r_1+r_2=3$}
There are ten expansion coefficients for rank-3 level.
\begin{subequations}
	\allowdisplaybreaks
	\begin{small}
		\bea
		&&\vec{\a} _{1,0,0}^{(3,0)}
		=\Bigg\{\frac{6 \left(4 f_{12} \left((D-1) M_1^2+(1-2 D) M_2^2\right)+3 D f_{12}^2-8 M_1^2 M_2^2\right)}{(D-1)
			(3 D-2) s_{11}}+\frac{12 (D-2) f_{12}}{(D-1) (3 D-2)},\nn
		&&\frac{6 D \left(-3
			f_{12}+2 M_1^2+4 M_2^2\right)}{(D-1) (3 D-2) s_{11}}-\frac{3 \left((4-8 D) f_{12} M_2^2+(3 D-2) f_{12}^2+8
			(D-1) M_1^2 M_2^2\right)}{(D-1) (3 D-2) s_{11}^2}\nn&&-\frac{12 (D-2)}{3 D^2-5 D+2},-\frac{6 M_1^2 \left((5 D-4)
			f_{12}+4 (1-2 D) M_2^2\right)}{(D-1) (3 D-2) s_{11}^2}-\frac{6 \left(3 D f_{12}+2 (D-2) M_1^2+4 (1-2 D)
			M_2^2\right)}{(D-1) (3 D-2) s_{11}}\nn&&-\frac{12 (D-2)}{3 D^2-5 D+2},\frac{6 \left((4 D-2) f_{12}+(D-2) M_1^2+4 (1-2 D) M_2^2\right)}{(D-1)
			(3 D-2) s_{11}^2}+\frac{6 (D-2)}{(D-1) (3 D-2) s_{11}},\nn&&\frac{12 \left(D \left(2 D^2-5 D+4\right) M_1^2-4
			\left(\left(D^2-3 D+1\right) M_2^2+(D-1)^2 M_3^2\right)-2 D f_{12}\right)}{(D-1) D \left(3 D^2-8 D+4\right)
			s_{11}}+\frac{12 \left(4 D^2-11 D+8\right)}{(D-1) \left(3 D^2-8 D+4\right)},\nn
		&&\frac{6 \left(D \left(3 D^2-2
			D-4\right) f_{12}-2 \left(D^2+2 D-4\right) M_1^2-4 D \left(2 D^2-4 D+1\right) M_2^2+4 (D-1) D
			M_3^2\right)}{(D-1) D \left(3 D^2-8 D+4\right) s_{11}}\nn&&+\frac{12 \left(D^2-6 D+6\right)}{(D-1) \left(3 D^2-8
			D+4\right)},-\frac{6 \left(3 D^2 f_{12}+2 (2 D-1) \left((D-2) M_1^2-2 D M_2^2\right)\right)}{(D-1) D (3
			D-2) s_{11}}-\frac{12 (D-2)}{3 D^2-5 D+2}\Bigg\}\co
		\eea
	\end{small}
	\begin{small}
		\bea
		&&\vec{\a} _{3,0,0}^{(3,0)}
		=\Bigg\{-\frac{2 (D+2) \left(4 f_{12} \left((D-1) M_1^2+(1-2 D) M_2^2\right)+3 D f_{12}^2-8 M_1^2
			M_2^2\right)}{(D-1) (3 D-2) s_{11}^2}-\frac{4 (D-2) (D+2) f_{12}}{(D-1) (3 D-2) s_{11}},\nn
		&&\frac{4 \left(2
			D^2-7 D+2\right) M_1^2+6 D (D+2) f_{12}-8 D (D+2) M_2^2}{(D-1) (3 D-2) s_{11}^2}+\frac{4 (D-2)
			(D+2)}{(D-1) (3 D-2) s_{11}}\nn&&+\frac{(D+2) \left(4 (1-2
			D) f_{12} M_2^2+(3 D-2) f_{12}^2+8 (D-1) M_1^2 M_2^2\right)}{(D-1) (3 D-2) s_{11}^3},\frac{4 (D-2) (D+2)}{(D-1) (3 D-2) s_{11}}\nn&&+\frac{2 (D+2) M_1^2 \left((5 D-4) f_{12}+4 (1-2 D) M_2^2\right)}{(D-1) (3 D-2)
			s_{11}^3}+\frac{2 (D+2) \left(3 D f_{12}+2 (D-2) M_1^2+4 (1-2 D) M_2^2\right)}{(D-1) (3 D-2)
			s_{11}^2},\nn&&\frac{8-2 D^2}{\left(3 D^2-5 D+2\right)
			s_{11}^2}-\frac{2 (D+2) \left((4 D-2) f_{12}+(D-2) M_1^2+4 (1-2 D) M_2^2\right)}{(D-1) (3 D-2)
			s_{11}^3},\nn&&\frac{4 \left(2 D^3-19 D^2+38 D-24\right)}{(D-1) \left(3 D^2-8 D+4\right) s_{11}}-\frac{4 (D+2)
			\left(D \left(2 D^2-5 D+4\right) M_1^2-2 D
			f_{12}\right)}{(D-1) D \left(3 D^2-8 D+4\right) s_{11}^2}\nn
		&&+\frac{16 (D+2)
			\left(\left(D^2-3 D+1\right) M_2^2+(D-1)^2 M_3^2\right)}{(D-1) D \left(3 D^2-8 D+4\right) s_{11}^2},-\frac{4 (D+2) \left(D^2-6 D+6\right)}{(D-1) \left(3 D^2-8 D+4\right)
			s_{11}}\nn&&-\frac{2 (D+2) \left(D \left(3 D^2-2 D-4\right)
			f_{12}-2 \left(D^2+2 D-4\right) M_1^2-4 D \left(2 D^2-4 D+1\right) M_2^2+4 (D-1) D M_3^2\right)}{(D-1) D
			\left(3 D^2-8 D+4\right) s_{11}^2},\nn&&\frac{2 (D+2) \left(3 D^2 f_{12}+2 (2 D-1) \left((D-2) M_1^2-2 D M_2^2\right)\right)}{(D-1) D (3
			D-2) s_{11}^2}+\frac{4 (D-2) (D+2)}{(D-1) (3 D-2) s_{11}}\Bigg\}\co
		\eea
	\end{small}
\begin{small}
\bea
&&\vec{\a} _{0,1,0}^{(2,1)}
=\Bigg\{\frac{4 D f_{12}}{3 D^2-5 D+2}+\frac{-4 f_{12} \left((D-1) M_1^2+(1-2 D) M_2^2\right)-2 f_{12}^2+8
	M_1^2 M_2^2}{(D-1) (3 D-2) s_{11}},-\frac{4 D}{3 D^2-5 D+2}\nn&&+\frac{2 M_2^2 \left((1-2 D) f_{12}+2 (D-1)
	M_1^2\right)}{(D-1) (3 D-2) s_{11}^2}+\frac{2 \left(2 (D-1) M_1^2-2 D M_2^2+f_{12}\right)}{(D-1) (3 D-2)
	s_{11}},-\frac{4 D}{3 D^2-5 D+2}\nn&&+\frac{2 M_1^2 \left((D-1) f_{12}+2 (1-2 D) M_2^2\right)}{(D-1) (3 D-2)
	s_{11}^2}+\frac{2 \left(4 (D-1) M_1^2+2 (1-2 D) M_2^2+f_{12}\right)}{(D-1) (3 D-2) s_{11}},\nn&&\frac{2
	D}{\left(3 D^2-5 D+2\right) s_{11}}-\frac{D f_{12}+4 (D-1) M_1^2+4 (1-2 D) M_2^2}{(D-1) (3 D-2)
	s_{11}^2},\frac{2 D (2 D-5)}{(D-1) \left(3 D^2-8
	D+4\right)}\nn&&+\frac{2 \left(D \left(-2 D^2+5 D-4\right) M_1^2+7 D^2 M_2^2+D^2 M_3^2+2 D f_{12}-20 D M_2^2+8
	M_2^2\right)}{(D-1) D \left(3 D^2-8 D+4\right) s_{11}},\nn&&\frac{-4 \left(D^2-5 D+4\right) M_1^2+4 D \left(2 D^2-4 D+1\right) M_2^2+2 (4-3 D) D f_{12}-4
	(D-1) D M_3^2}{(D-1) D \left(3 D^2-8 D+4\right) s_{11}}\nn&&+\frac{4 \left(D^2-D-1\right)}{(D-1) \left(3 D^2-8
	D+4\right)},\frac{2 \left(2 (D-1) M_1^2+2 (1-2 D) M_2^2+f_{12}\right)}{(D-1) (3 D-2) s_{11}}-\frac{4 D}{3
	D^2-5 D+2}\Bigg\}\co
\eea
\end{small}
\begin{small}
\bea
&&\vec{\a} _{1,0,1}^{(2,1)}
=\Bigg\{\frac{16 f_{12}}{2-3 D}-\frac{4 \left(2 f_{12} \left((D-1) M_1^2+(1-2 D) M_2^2\right)+f_{12}^2-4 M_1^2
	M_2^2\right)}{(D-1) (3 D-2) s_{11}},\frac{16}{3
	D-2}\nn
&&+\frac{4 M_2^2 \left((1-2 D) f_{12}+2 (D-1) M_1^2\right)}{(D-1) (3 D-2)
	s_{11}^2}+\frac{4 \left(2 (D-1) M_1^2-2 D M_2^2+f_{12}\right)}{(D-1) (3 D-2) s_{11}},\frac{4 M_1^2 \left((D-1) f_{12}+2 (1-2 D) M_2^2\right)}{(D-1) (3 D-2) s_{11}^2}\nn&&+\frac{16}{3 D-2}+\frac{4 \left(4 (D-1)
	M_1^2+2 (1-2 D) M_2^2+f_{12}\right)}{(D-1) (3 D-2) s_{11}},\frac{8}{(2-3 D)
	s_{11}}-\frac{2 \left(D f_{12}+4 (D-1) M_1^2+4 (1-2 D) M_2^2\right)}{(D-1) (3 D-2) s_{11}^2},\nn&&\frac{4
	\left(\left(3 D^2-8 D+6\right) D f_{12}-3 D^3 M_2^2+3 D^3 M_3^2+15 D^2 M_2^2-7 D^2 M_3^2+\left(-5 D^2+13
	D-8\right) D M_1^2\right)}{(D-1) D \left(3 D^2-8 D+4\right)
	s_{11}}\nn&&+\frac{-24 D M_2^2+4 D M_3^2+8 M_2^2}{(D-1) D \left(3 D^2-8 D+4\right)
s_{11}}+\frac{48-28 D}{3 D^2-8 D+4},\frac{40-16 D}{3 D^2-8
D+4}+\frac{4 (4-3 D) D f_{12}-8 (D-1) D M_3^2}{(D-1) D \left(3 D^2-8 D+4\right) s_{11}}\nn&&+\frac{-8 \left(D^2-5 D+4\right) M_1^2+8 D \left(2 D^2-4 D+1\right)
	M_2^2}{(D-1) D \left(3 D^2-8 D+4\right) s_{11}},\frac{4 \left(2 (D-1) M_1^2+2 (1-2 D) M_2^2+f_{12}\right)}{(D-1) (3 D-2) s_{11}}+\frac{16}{3
	D-2}\Bigg\}\co
\eea
\end{small}
\begin{small}
\bea
&&\vec{\a} _{2,1,0}^{(2,1)}
=\Bigg\{\frac{2 (D+2) \left(2 f_{12} \left((D-1) M_1^2+(1-2 D) M_2^2\right)+f_{12}^2-4 M_1^2
	M_2^2\right)}{(D-1) (3 D-2) s_{11}^2}-\frac{4 (D-2)^2 f_{12}}{(D-1) (3 D-2) s_{11}},\nn&&-\frac{2 (D+2) \left(2
	(D-1) M_1^2-2 D M_2^2+f_{12}\right)}{(D-1) (3 D-2) s_{11}^2}+\frac{2 (D+2) M_2^2 \left((2 D-1) f_{12}-2
	(D-1) M_1^2\right)}{(D-1) (3 D-2) s_{11}^3}\nn&&+\frac{4 (D-2)^2}{(D-1) (3 D-2) s_{11}},\frac{4 \left(D^2-7
	D+6\right) M_1^2+4 \left(2 D^2+3 D-2\right) M_2^2-2 (D+2) f_{12}}{(D-1) (3 D-2) s_{11}^2}\nn&&-\frac{2 (D+2)
	M_1^2 \left((D-1) f_{12}+2 (1-2 D) M_2^2\right)}{(D-1) (3 D-2) s_{11}^3}+\frac{4 (D-2)^2}{(D-1) (3 D-2)
	s_{11}},-\frac{2
	(D-2)^2}{(D-1) (3 D-2) s_{11}^2}\nn&&+\frac{(D+2) \left(D f_{12}+4 (D-1) M_1^2+4 (1-2 D) M_2^2\right)}{(D-1) (3 D-2) s_{11}^3},\frac{2 \left(-2 \left(3 D^2-7 D+6\right) D f_{12}-D^3
	M_2^2\right)}{(D-1) D \left(3 D^2-8 D+4\right) s_{11}^2}\nn&&+\frac{2 \left(-7 D^3 M_3^2-10 D^2 M_2^2+14 D^2 M_3^2\right)}{(D-1) D \left(3 D^2-8 D+4\right) s_{11}^2}+\frac{-4 D^3+38 D^2-76
	D+48}{(D-1) \left(3 D^2-8 D+4\right) s_{11}}+\frac{2 \left(40 D
	M_2^2-8 D M_3^2-16 M_2^2\right)}{(D-1) D \left(3 D^2-8 D+4\right) s_{11}^2}\nn&&+\frac{2 \left(2 D^3+5 D^2-22 D+16\right) D M_1^2}{(D-1) D \left(3 D^2-8 D+4\right) s_{11}^2},-\frac{2 (D+2) \left(-2 \left(D^2-5 D+4\right) M_1^2+2 D
	\left(2 D^2-4 D+1\right) M_2^2\right)}{(D-1) D \left(3 D^2-8 D+4\right)
	s_{11}^2}\nn&&-\frac{2 (D+2) \left((4-3 D) D f_{12}-2 (D-1) D M_3^2\right)}{(D-1) D \left(3 D^2-8 D+4\right)
	s_{11}^2}-\frac{4 \left(D^3-5 D^2+13 D-10\right)}{(D-1) \left(3 D^2-8 D+4\right) s_{11}},\frac{4
	(D-2)^2}{(D-1) (3 D-2) s_{11}}\nn&&-\frac{2 (D+2) \left(2 (D-1) M_1^2+2 (1-2 D) M_2^2+f_{12}\right)}{(D-1) (3
	D-2) s_{11}^2}\Bigg\}\co
\eea
\end{small}
\begin{small}
\bea
&&\vec{\a} _{0,1,0}^{(0,3)}
=\Bigg\{\frac{6 \left((4-8 D) f_{12} M_1^2+4 (D-1) f_{12} M_2^2+3 D f_{12}^2-8 M_1^2 M_2^2\right)}{(D-1) (3
	D-2) s_{11}}+\frac{12 (D-2) f_{12}}{(D-1) (3 D-2)},-\frac{12 (D-2)}{3 D^2-5 D+2}\nn&&-\frac{6 M_2^2 \left((5
	D-4) f_{12}+4 (1-2 D) M_1^2\right)}{(D-1) (3 D-2) s_{11}^2}-\frac{6 \left(3 D f_{12}+(4-8 D) M_1^2+2 (D-2)
	M_2^2\right)}{(D-1) (3 D-2) s_{11}},-\frac{12 (D-2)}{3 D^2-5 D+2}\nn&&+\frac{6 D \left(-3 f_{12}+4 M_1^2+2
	M_2^2\right)}{(D-1) (3 D-2) s_{11}}-\frac{3 \left((4-8 D) f_{12} M_1^2+(3 D-2) f_{12}^2+8 (D-1) M_1^2
	M_2^2\right)}{(D-1) (3 D-2) s_{11}^2},\nn&&\frac{6 \left((4 D-2) f_{12}+(4-8 D) M_1^2+(D-2) M_2^2\right)}{(D-1)
	(3 D-2) s_{11}^2}+\frac{6 (D-2)}{(D-1) (3 D-2) s_{11}},\frac{12 \left(D^2-6 D+6\right)}{(D-1) \left(3 D^2-8 D+4\right)}\nn&&+\frac{6 \left(D \left(3 D^2-2 D-4\right) f_{12}-4 D
	\left(2 D^2-4 D+1\right) M_1^2-2 \left(D^2+2 D-4\right) M_2^2+4 (D-1) D M_3^2\right)}{(D-1) D \left(3 D^2-8
	D+4\right) s_{11}},\nn&&\frac{12 \left(2 D^3
	M_2^2-5 D^2 M_2^2-4 D^2 M_3^2-4 \left(D^2-3 D+1\right) M_1^2-2 D f_{12}+4 D M_2^2+8 D M_3^2-4
	M_3^2\right)}{(D-1) D \left(3 D^2-8 D+4\right) s_{11}}\nn&&+\frac{12 \left(4 D^2-11 D+8\right)}{(D-1) \left(3
	D^2-8 D+4\right)},-\frac{6 \left(3 D^2 f_{12}+4 (1-2 D) D M_1^2+2 (D-2) (2 D-1) M_2^2\right)}{(D-1) D (3
	D-2) s_{11}}-\frac{12 (D-2)}{3 D^2-5 D+2}\Bigg\}\co\nn~~~
\eea
\end{small}
\begin{small}
\bea
&&\vec{\a} _{0,3,0}^{(0,3)}
=\Bigg\{-\frac{4 (D-2) (D+2) f_{12}}{(D-1) (3 D-2) s_{11}}-\frac{2 (D+2) \left(3 D f_{12}^2+(4-8 D) M_1^2
	f_{12}+4 (D-1) M_2^2 f_{12}-8 M_1^2 M_2^2\right)}{(D-1) (3 D-2) s_{11}^2},\nn&&\frac{2 (D+2) \left(4 (1-2 D)
	M_1^2+(5 D-4) f_{12}\right) M_2^2}{(D-1) (3 D-2) s_{11}^3}+\frac{2 (D+2) \left((4-8 D) M_1^2+2 (D-2) M_2^2+3 D f_{12}\right)}{(D-1) (3 D-2) s_{11}^2}\nn&&+\frac{4 (D-2) (D+2)}{(D-1) (3 D-2)
	s_{11}},\frac{4
	\left(D^2-4\right)}{(D-1) (3 D-2) s_{11}}+\frac{-8 D (D+2) M_1^2+4 \left(2 D^2-7 D+2\right) M_2^2+6 D (D+2)
	f_{12}}{(D-1) (3 D-2) s_{11}^2}\nn&&+\frac{(D+2) \left((3 D-2) f_{12}^2+4 (1-2 D) M_1^2 f_{12}+8 (D-1) M_1^2
	M_2^2\right)}{(D-1) (3 D-2) s_{11}^3},-\frac{2 \left(D^2-4\right)}{\left(3 D^2-5 D+2\right)
	s_{11}^2}\nn&&-\frac{2 (D+2) \left((4-8 D) M_1^2+(D-2) M_2^2+(4 D-2) f_{12}\right)}{(D-1) (3 D-2)
	s_{11}^3},-\frac{4 (D+2) \left(D^2-6 D+6\right)}{(D-1) \left(3 D^2-8 D+4\right) s_{11}}\nn&&-\frac{2 (D+2)
	\left(-4 D \left(2 D^2-4 D+1\right) M_1^2-2 \left(D^2+2 D-4\right) M_2^2+4 (D-1) D M_3^2+D \left(3 D^2-2
	D-4\right) f_{12}\right)}{(D-1) D \left(3 D^2-8 D+4\right) s_{11}^2},\nn&&\frac{4 \left(2 D^3-19 D^2+38
	D-24\right)}{(D-1) \left(3 D^2-8 D+4\right) s_{11}}-\frac{4 (D+2) \left(2 M_2^2 D^3-5 M_2^2 D^2-4 M_3^2
	D^2+4 M_2^2 D+8 M_3^2 D-2D f_{12} \right)}{(D-1) D \left(3 D^2-8
	D+4\right) s_{11}^2}\nn&&-\frac{4 (D+2) \left(-4 \left(D^2-3 D+1\right) M_1^2-4 M_3^2\right)}{(D-1) D \left(3 D^2-8
	D+4\right) s_{11}^2},\frac{2 \left(3 f_{12} D^2+4 (1-2 D) M_1^2
	D+2 (D-2) (2 D-1) M_2^2\right) (D+2)}{(D-1) D (3 D-2) s_{11}^2}\nn&&+\frac{4 (D-2) (D+2)}{(D-1) (3 D-2) s_{11}}\Bigg\}\co
\eea
\end{small}
\begin{small}
\bea
&&\vec{\a} _{1,0,0}^{(1,2)}
=\bigg\{\frac{4 D f_{12}}{3 D^2-5 D+2}+\frac{4 f_{12} \left((2 D-1) M_1^2-(D-1) M_2^2\right)-2 f_{12}^2+8 M_1^2
	M_2^2}{(D-1) (3 D-2) s_{11}},-\frac{4 D}{3 D^2-5 D+2}\nn&&+\frac{2 M_2^2 \left((D-1) f_{12}+2 (1-2 D)
	M_1^2\right)}{(D-1) (3 D-2) s_{11}^2}+\frac{2 \left((2-4 D) M_1^2+4 (D-1) M_2^2+f_{12}\right)}{(D-1) (3
	D-2) s_{11}},-\frac{4 D}{3 D^2-5 D+2}\nn&&+\frac{2 M_1^2 \left((1-2 D) f_{12}+2 (D-1) M_2^2\right)}{(D-1) (3
	D-2) s_{11}^2}+\frac{2 \left(-2 D M_1^2+2 (D-1) M_2^2+f_{12}\right)}{(D-1) (3 D-2) s_{11}},\frac{2
	D}{\left(3 D^2-5 D+2\right) s_{11}}\nn&&+\frac{-D f_{12}+(8 D-4) M_1^2-4 (D-1) M_2^2}{(D-1) (3 D-2)
	s_{11}^2},\frac{4 \left(D^2-D-1\right)}{(D-1) \left(3 D^2-8
	D+4\right)}+\frac{4 D \left(2 D^2-4 D+1\right) M_1^2}{(D-1) D \left(3 D^2-8 D+4\right) s_{11}}\nn&&+\frac{2 (4-3 D) D f_{12}-4 (D-1) \left((D-4) M_2^2+D
	M_3^2\right)}{(D-1) D \left(3 D^2-8 D+4\right) s_{11}},\frac{2 D (2 D-5)}{(D-1) \left(3
	D^2-8 D+4\right)}\nn&&+\frac{2 \left(\left(7 D^2-20 D+8\right) M_1^2+D \left(\left(-2 D^2+5 D-4\right) M_2^2+D
	M_3^2\right)+2 D f_{12}\right)}{(D-1) D \left(3 D^2-8 D+4\right) s_{11}},-\frac{4
	D}{3 D^2-5 D+2}\nn&&+\frac{2 \left(2 (1-2 D) M_1^2+2 (D-1) M_2^2+f_{12}\right)}{(D-1) (3 D-2) s_{11}}\bigg\}\co
\eea
\end{small}
\begin{small}
\bea
&&\vec{\a} _{0,1,1}^{(1,2)}=\bigg\{\frac{8 f_{12} \left((2 D-1) M_1^2-(D-1) M_2^2\right)-4 f_{12}^2+16 M_1^2 M_2^2}{(D-1) (3 D-2)
	s_{11}}+\frac{16 f_{12}}{2-3 D},\frac{4 M_2^2 \left((D-1) f_{12}+2 (1-2 D) M_1^2\right)}{(D-1) (3 D-2)
	s_{11}^2}\nn&&+\frac{4 \left((2-4 D) M_1^2+4 (D-1) M_2^2+f_{12}\right)}{(D-1) (3 D-2) s_{11}}+\frac{16}{3
	D-2},\frac{4 M_1^2 \left((1-2 D) f_{12}+2 (D-1) M_2^2\right)}{(D-1) (3 D-2) s_{11}^2}\nn&&+\frac{4 \left(-2 D
	M_1^2+2 (D-1) M_2^2+f_{12}\right)}{(D-1) (3 D-2) s_{11}}+\frac{16}{3 D-2},\frac{8}{(2-3 D) s_{11}}-\frac{2
	\left(D f_{12}+(4-8 D) M_1^2+4 (D-1) M_2^2\right)}{(D-1) (3 D-2) s_{11}^2},\nn&&\frac{8 D \left(2 D^2-4
	D+1\right) M_1^2+4 (4-3 D) D f_{12}-8 (D-1) \left((D-4) M_2^2+D M_3^2\right)}{(D-1) D \left(3 D^2-8
	D+4\right) s_{11}}+\frac{40-16 D}{3 D^2-8 D+4},\nn&&\frac{4 \left(D \left(3 D^2-8 D+6\right) f_{12}+\left(-3
	D^3+15 D^2-24 D+8\right) M_1^2-(D-1) D \left((5 D-8) M_2^2+(4-3 D) M_3^2\right)\right)}{(D-1) D \left(3
	D^2-8 D+4\right) s_{11}}\nn&&+\frac{48-28 D}{3 D^2-8 D+4},\frac{4 \left(2 (1-2 D) M_1^2+2 (D-1)
	M_2^2+f_{12}\right)}{(D-1) (3 D-2) s_{11}}+\frac{16}{3 D-2}\bigg\}\co
\eea
\end{small}
\begin{small}
\bea
&&\vec{\a} _{1,2,0}^{(1,2)}
=\bigg\{\frac{2 (D+2) \left(-2 f_{12} \left((2 D-1) M_1^2-(D-1) M_2^2\right)+f_{12}^2-4 M_1^2
	M_2^2\right)}{(D-1) (3 D-2) s_{11}^2}-\frac{4 (D-2)^2 f_{12}}{(D-1) (3 D-2) s_{11}},\nn&&\frac{4 \left(2 D^2+3
	D-2\right) M_1^2+4 \left(D^2-7 D+6\right) M_2^2-2 (D+2) f_{12}}{(D-1) (3 D-2) s_{11}^2}-\frac{2 (D+2) M_2^2
	\left((D-1) f_{12}+2 (1-2 D) M_1^2\right)}{(D-1) (3 D-2) s_{11}^3}\nn&&+\frac{4 (D-2)^2}{(D-1) (3 D-2)
	s_{11}},-\frac{2 (D+2) \left(-2 D M_1^2+2 (D-1) M_2^2+f_{12}\right)}{(D-1) (3 D-2) s_{11}^2}+\frac{4 (D-2)^2}{(D-1) (3 D-2)
	s_{11}}\nn&&+\frac{2 (D+2)
	M_1^2 \left((2 D-1) f_{12}-2 (D-1) M_2^2\right)}{(D-1) (3 D-2) s_{11}^3},\frac{(D+2) \left(D f_{12}+4 (1-2 D) M_1^2+4 (D-1) M_2^2\right)}{(D-1) (3 D-2) s_{11}^3}\nn&&-\frac{2
	(D-2)^2}{(D-1) (3 D-2) s_{11}^2},-\frac{4
	\left(D^3-5 D^2+13 D-10\right)}{(D-1) \left(3 D^2-8 D+4\right) s_{11}}\nn&&-\frac{2 (D+2) \left(2 D \left(2 D^2-4 D+1\right) M_1^2+(4-3 D) D f_{12}-2
	(D-1) \left((D-4) M_2^2+D M_3^2\right)\right)}{(D-1) D \left(3 D^2-8 D+4\right) s_{11}^2},\nn&&\frac{-4 D^3+38 D^2-76 D+48}{(D-1)
	\left(3 D^2-8 D+4\right) s_{11}}-\frac{2 \left(D \left(7 D^2-14 D+8\right) M_3^2\right)}{(D-1)
	D \left(3 D^2-8 D+4\right) s_{11}^2}
\nn&&-\frac{2 \left(2 D \left(3 D^2-7 D+6\right) f_{12}+\left(D^3+10 D^2-40
	D+16\right) M_1^2-D \left(2 D^3+5 D^2-22 D+16\right) M_2^2\right)}{(D-1)
	D \left(3 D^2-8 D+4\right) s_{11}^2}\nn&&,\frac{4 (D-2)^2}{(D-1) (3 D-2) s_{11}}-\frac{2 (D+2) \left(2 (1-2 D)
	M_1^2+2 (D-1) M_2^2+f_{12}\right)}{(D-1) (3 D-2) s_{11}^2}\bigg\}\ed~~~
\eea
\end{small}
\end{subequations}

\bibliographystyle{JHEP}

\bibliography{reference}
	
\end{document}